\documentclass[twocolumn, journal]{IEEEtran}
\IEEEoverridecommandlockouts
\usepackage{color}
\usepackage{graphicx}
\usepackage{epstopdf}
\usepackage{amsmath}
\usepackage{amssymb}
\usepackage{algorithm}
\usepackage{algorithmic}
\usepackage{amsmath}
\usepackage{multirow}
\usepackage{booktabs}
\usepackage{array}
\usepackage{amsthm}
\usepackage{stfloats}
\usepackage{caption}
\usepackage{subfigure}
\usepackage{bm}
\usepackage{setspace}
\usepackage{diagbox}
\usepackage{enumitem}

\newcommand{\be}{\begin{equation}}
\newcommand{\ee}{\end{equation}}
\newcommand{\bea}{\begin{eqnarray}}
\newcommand{\eea}{\end{eqnarray}}
\newcommand{\ba}{\begin{array}}
\newcommand{\ea}{\end{array}}

\allowdisplaybreaks[4]
\usepackage{flushend}

\def\BibTeX{{\rm B\kern-.05em{\sc i\kern-.025em b}\kern-.08em
    T\kern-.1667em\lower.7ex\hbox{E}\kern-.125emX}}
\begin{document}

\title{Distributed Distortion-Aware Beamforming Designs for Cell-Free mMIMO Systems
\thanks{M. Liu and M. Li are with the School of Information and Communication Engineering, Dalian University of Technology, Dalian 116024, China (e-mail: liumengzhen@mail.dlut.edu.cn, mli@dlut.edu.cn).}
\thanks{R. Liu was with the School of Information and Communication Engineering, Dalian University of Technology, Dalian 116024, China. She is currently with the Center for Pervasive Communications and Computing, University of California, Irvine, CA 92697, USA (e-mail: rangl2@uci.edu).}
\thanks{Q. Liu is with the School of Computer Science and Technology, Dalian University of Technology, Dalian 116024, China (e-mail: qianliu@dlut.edu.cn).}}

\author{Mengzhen Liu,~\IEEEmembership{Student Member,~IEEE,}
        Ming Li,~\IEEEmembership{Senior Member,~IEEE,}
        Rang Liu,~\IEEEmembership{Member,~IEEE,}\\
        and Qian Liu,~\IEEEmembership{Member,~IEEE}
}

\maketitle
\pagestyle{empty}  
\thispagestyle{empty} 

\begin{abstract}
Cell-free massive multi-input multi-output (CF-mMIMO) systems have emerged as a promising paradigm for next-generation wireless communications, offering enhanced spectral efficiency and coverage through distributed antenna arrays. However, the non-linearity of power amplifiers (PAs) in these arrays introduce spatial distortion, which may significantly degrade system performance. This paper presents the first investigation of distortion-aware beamforming in a distributed framework tailored for CF-mMIMO systems, enabling pre-compensation for beam dispersion caused by nonlinear PA distortion. Using a third-order memoryless polynomial distortion model, the impact of the nonlinear PA on the performance of CF-mMIMO systems is firstly analyzed by evaluating the signal-to-interference-noise-and-distortion ratio (SINDR) at user equipment (UE). Then, we develop two distributed distortion-aware beamforming designs based on ring topology and star topology, respectively. In particular, the ring-topology-based fully-distributed approach reduces interconnection costs and computational complexity, while the star-topology-based partially-distributed scheme leverages the superior computation capability of the central processor to achieve improved sum-rate performance. Extensive simulations demonstrate the effectiveness of the proposed distortion-aware beamforming designs in mitigating the effect of nonlinear PA distortion, while also reducing computational complexity and backhaul information exchange in CF-mMIMO systems.
\end{abstract}


\section{Introduction}
With the rapid expansion of wireless communication networks, the number of mobile devices is experiencing exponential growth, driving increased demands for higher data rates and broader coverage. To meet these escalating requirements, cell-free massive multiple-input multiple-output (CF-mMIMO) has emerged as a promising technology for future wireless systems. Unlike conventional cellular networks, CF-mMIMO enables simultaneous service to multiple user equipments (UEs) through a network of distributed multi-antenna base stations (BSs), or access points (APs), interconnected with a central network controller \cite{E. Bjornson 2019 MIMO}.

As a type of distributed extremely large-scale antenna array (ELAA) system, CF-mMIMO system essentially leverages high spatial diversity and beamforming gains, which are critically dependent on the cooperative beamforming of multiple BSs. However, in a practical cost-efficient CF-mMIMO systems, which involves the use of low-quality components for a large number of antenna elements, hardware impairments and deficiencies are inevitable.
These impairments and deficiencies degrade beamforming performance, posing significant challenges to maintaining reliable communication quality \cite{E. Bjornson 2014}-\cite{E. Jorswieck 2023}.
Specifically, major hardware limitations include unsynchronized phase shifts \cite{E. G. Larsson OJCS 2024}, channel miscalibration \cite{X. Li PIMRC 2023}, low-precision analog-to-digital converters (ADCs) \cite{A. L. Swindlehurst TWC 2021}, and power amplifier non-linearities \cite{C. Mollen 2018}, \cite{N. N. Moghadam 2018}.
Therefore, future CF-mMIMO systems must move beyond ideal hardware assumptions and develop robust beamforming algorithms capable of adapting to and compensating for these non-idealities.

Among these hardware  impairments, the nonlinear distortion induced by imperfect power amplifiers (PAs) is a key factor  which degrades the beamforming performance of CF-mMIMO systems. In particular, due to the limited range of linear amplification, when the input beamformed signal exhibits large amplitude variations, the PA will not only perform nonlinear amplification to the signal magnitude, but also introduce distortion to the signal phase. The combined effects of nonlinear PAs on the amplitude and phase of signals will generate substantial spatial distortion beams toward various unintended directions and exacerbate the interference among UEs. Undoubtedly, this beam dispersion and intensified interference will critically affect the reception of useful signals, leading to severe performance degradation. Therefore, to mitigate these adverse effects of nonlinear PAs, it is essential to explore distortion-aware beamforming techniques that enhance the robustness of beamforming and improve overall system reliability.

Given the greatly negative impact of nonlinear PAs on communication quality, researchers have recently begun to explore solutions to mitigate this effect. The authors in \cite{E. Bjornson 2024} analytically investigated the angular directions and depth of the nonlinear distortion in both near- and far-field channels. Besides, the correlation of the distortion signals across different antennas and its influence on system spectral efficiency have been sufficiently illustrated in \cite{E. Bjornson 2019}. In addition to analyzing the characteristics of distortion beams and assessing their impact on network performance, several studies have explored robust beamforming designs to mitigate distortion and maintain service quality. For instance, \cite{B. Liu 2024} derived an optimal phase shift matrix for a transparent amplifying intelligent surface to maximize uplink spectral efficiency under nonlinear distortion.  Moreover, distortion-aware linear precoder designs were developed using the projected gradient ascent method in \cite{S. R. Aghdam 2019} and \cite{M. Wu 2022}.

Existing  distortion-aware beamforming studies primarily focus on single-BS scenarios. However, in CF-mMIMO scenarios where multiple BSs simultaneously serve UEs, applying these strategies to pre-compensate for nonlinear distortion becomes impractical  \cite{JSTSP}, \cite{W. Yu decentralized 2024}.  Specifically, effective coordination among BSs to manage increased distortion beams and stronger interference requires the cloud processor to aggregate and process information from all BSs and UEs. This leads to  an exponential increase of the computational burden on the central processor and the information transmission load on the backhaul link \cite{T.-H. Chang ICC 2023}.
To address these challenges, particularly the computational limitations at the central processor and the bandwidth constraints of the backhaul link, it is crucial to employ distributed signal processing techniques for implementing distortion-aware beamforming \cite{T.-H. Chang ICASSP 2023}-\cite{E. G. Larsson WCL-2 2024}.

It is noteworthy that the distributed beamforming designs are naturally well-suited for cell-free networks, as multiple BSs inherently form a distributed system, each equipped with local computational capabilities and cost-effective baseband signal processing units. By allowing each BS to collect only its relevant channel information and perform parallel computations, distributed signal processing schemes can substantially reduce backhaul overhead and alleviate the computational burden on the central processor \cite{E. Jorswieck 2016}-\cite{H. Vincent Poor 2021}. Additionally, distributing computational tasks enhance the network's scalability and reliability, enabling support for a large number of devices and providing fault tolerance at individual nodes.

In light of these benefits offered by distributed signal processing, several studies have explored approaches where multiple entities collaboratively address the global optimization problem by solving local sub-problems. In \cite{A. Tolli 2011}-\cite{T.-H. Chang J-2 2020}, dual decomposition methods were applied to manage the coupled interference terms between BSs in multi-cell networks.  Additionally, the global energy efficiency optimization problem was decomposed using the alternating direction method of multipliers (ADMM), which is also known as the global consensus problem \cite{ADMM 2011}-\cite{F. Han ADMM 2017}. The authors in \cite{P. Ni decentralized 2023} employed weighted minimize mean squared error (WMMSE) to transform the weighted sum-rate maximization problem into an equivalent weighted sum-mean-square-error (sum-MSE) minimization problem,  where the sub-problems for each BS were fully decoupled. Moreover, distributed learning-based decomposition techniques have also been proposed for similar purposes \cite{H. Zhang decentralized 2023}, \cite{Z. Wang decentralized 2022}. However, to the best of the authors' knowledge, no existing research has investigated the application of distributed beamforming designs in CF-mMIMO systems to mitigate nonlinear distortion effect.

Motivated by the above discussions, this paper investigates distributed distortion-aware beamforming designs for CF-mMIMO systems under both ring and star topologies. The main contributions are summarized as follows:
\begin{itemize}
  \item
   Utilizing a typical third-order memoryless polynomial distortion model of the nonlinear PAs, we provide a comprehensive analysis of the impact of nonlinear PA amplification on CF-mMIMO system performance. To mitigate these nonlinear distortion effects, we present the first investigation of distortion-aware beamforming designs in a distributed framework tailored for CF-mMIMO systems, with the objective of maximizing the sum-rate subject to the transmit power constraints of each BS.

  \item We first introduce a ring-topology-based fully-distributed distortion-aware beamforming design algorithm, aiming to alleviate the computational load on the central processor. To be specific, each local BS sequentially computes its own beamforming, updates its relevant information, and relays the global data to the next neighboring BS until convergence is achieved. Within this framework, we develop a penalty-majorization-minimization-based (penalty-MM-based) distributed beamforming design algorithm to decompose the high-order terms in the original highly non-convex objective function, and alternately solve the corresponding sub-problems.

  \item To leverage the central processor's superior computation resources for achieving performance improvement, a star-topology-based partially-distributed distortion-aware beamforming design algorithm is also presented. In this framework, each BS concurrently executes local beamforming design and then uploads its calculated results to the central processor for information aggregation. In addition to utilizing penalty-MM-based algorithm to deal with the challenges caused by nonlinear distortion, we also develop a consensus-ADMM-based algorithm to ensure information consistency between the local BSs and the central processor.

  \item Extensive simulations are conducted to verify the effectiveness of the proposed distributed distortion-aware beamforming designs in mitigating nonlinear distortion effect. Additionally, these simulations demonstrate the capability of the proposed designs to reduce computational complexity and backhaul signaling overhead in CF-mMIMO systems.
\end{itemize}

\textit{Notations}: Boldface lower-case and upper-case letters indicate column vectors and matrices, respectively. $(\cdot)^{*}$, $(\cdot)^{T}$, and $(\cdot)^{H}$ denote the conjugate, transpose, and transpose-conjugate, respectively. $|a|$, $\|\mathbf{a}\|_{2}$, and $\|\mathbf{A}\|_{F}$ are the magnitude of scalar $a$, the norm of vector $\mathbf{a}$, and the Frobenius norm of matrix $\mathbf{A}$. $|\mathbf{A}|$ represents the element-wise magnitude of matrix $\mathbf{A}$. $\mathrm{Tr}(\mathbf{A})$ denotes the trace of matrix $\mathbf{A}$. The vectorization operation $\mathrm{vec}(\mathbf{A})$ stacks the $\mathbf{A}$'s columns into a long column vector. Notations $\odot$ and $\otimes$ are the Hadamard product and Kronecker product of matrices, respectively. $\mathrm{diag}\{\mathbf{a}\}$ indicates the diagonal matrix whose diagonals are the elements of $\mathbf{a}$, while $\mathrm{diag}\{\mathbf{A}\}$ is a diagonal matrix whose elements are the main diagonals of $\mathbf{A}$. $\mathbf{1}_{M}$, $\mathbf{1}_{M\times M}$, and $\mathbf{I}_{M}$ represent a $M\times 1$ vector of ones, a $M\times M$ matrix of ones, and an $M\times M$ identity matrix, respectively. $\Re\{\cdot\}$ denotes the real part of a complex number. The statistical expectation is given by $\mathbb{E}\{\cdot\}$. Finally, $\mathbf{A}(i,:)$ and $\mathbf{A}(i,j)$ denote the $i$-th row and $(i,j)$-th element of matrix $\mathbf{A}$, respectively.

\begin{figure}[!t]
  \centering
  \vspace{-0.2 cm}
  \includegraphics[width= 2.7 in]{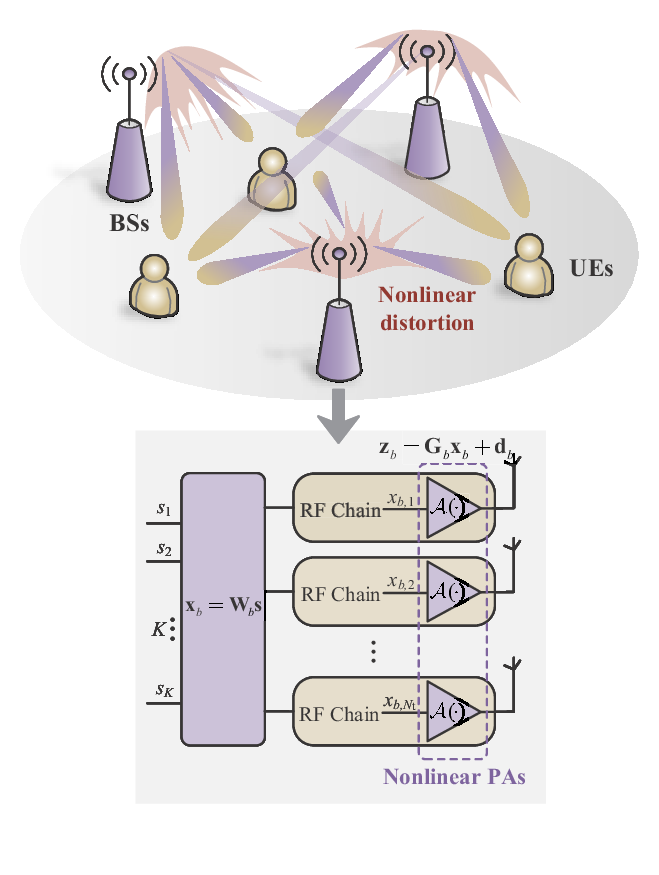}
 \vspace{-0.7cm}
 \caption{The architecture of CF-mMIMO system and the illustration of nonlinear distortion effect.}
  \label{fig:systemmodel}
  \vspace{-0.4 cm}
\end{figure}

\section{System Model and Problem Formulation}

\subsection{System Model}
We consider a CF-mMIMO system, as shown in Fig. \ref{fig:systemmodel}. Each BS is equipped with a uniform linear array (ULA) where each antenna is connected to its own radio frequency (RF) chain and PA. By employing suitable beamforming techniques, all BSs collaboratively serve multiple single-antenna UEs over the same frequency-time resources.
Specifically, the considered CF-mMIMO system has $B$ multi-antenna BSs and $K$ single-antenna UEs. Each BS is equipped with $N_{\mathrm{t}}$ antennas.
Let $\mathcal{B}\triangleq \{1, 2, \ldots, B\}$, $\mathcal{N}\triangleq \{1, 2, \ldots, N_{\mathrm{t}}\}$, $\mathcal{K}\triangleq \{1, 2, \ldots, K\}$ denote the sets of BSs, BS antennas, and UEs, respectively. Define $\mathbf{s}\triangleq[s_{1}, s_{2}, \ldots, s_{K}]^{T}\in \mathbb{C}^{K}$ as the transmitted symbols satisfying $\mathbb{E}\{\mathbf{s}\mathbf{s}^{H}\}=\mathbf{I}_{K}$. With a linear beamforming matrix $\mathbf{W}_{b}\triangleq[\mathbf{w}_{b,1}, \mathbf{w}_{b,2}, \ldots, \mathbf{w}_{b,K}]\in \mathbb{C}^{N_{\mathrm{t}}\times K}$ at the $b$-th BS, the beamformed signal is obtained by $\mathbf{x}_{b}=\mathbf{W}_{b}\mathbf{s}\in \mathbb{C}^{N_{\mathrm{t}}}$.
Then, the baseband signal $\mathbf{x}_{b}$ will be up-converted to carrier frequency through RF chains and amplified via PAs before being emitted from $N_{\mathrm{t}}$ antennas.

While beamforming design has been extensively investigated over the past decades, most existing studies assume perfect linear amplification, wherein signal amplitudes remain within the linear operating range of PAs. Nevertheless, the linear amplification regime is often impractical in real-world applications due to hardware limitations. Nonlinear amplification inevitably arises in the cost-efficient implementation of CF-mMIMO systems, particularly when low-quality components are used in densely deployed antenna arrays.
Consequently, the nonlinear amplification of beamformed signals to be transmitted via multiple antennas will cause beam distortion,  manifesting as unpredictable sidelobes in undesired directions. This beam distortion has a substantial adverse effect on the performance of CF-mMIMO systems, where the dense deployment of BSs and UEs makes beamforming based interference management essential.

To explicitly analyze the impact of beam distortion, we adopt a widely-used third-order memoryless polynomial distortion model to describe the non-linearity of PAs at the transmitter \cite{E. Bjornson 2024}. We assume that all the BSs in this cell-free network have identical amplifier models and parameters, resulting in the same nonlinear distortion behavior across all BSs. Thus, the nonlinear distorted signal at the $n$-th antenna of the $b$-th BS can be expressed by
\begin{equation}
\label{eq:distortion_model1}
z_{b,n}=\mathcal{A}(x_{b,n})=\beta_{1}x_{b,n}+\beta_{3}x_{b,n}|x_{b,n}|^{2}, ~\forall b, n,
\end{equation}
where $\mathcal{A}(\cdot)$ denotes the nonlinear operation of the PA, $\beta_{1}=1, \beta_{3}\in \mathbb{C}$ are the nonlinear model parameters for the first- and third- order terms, respectively. The parameter $\beta_{3}$ typically takes complex values, capturing both amplitude-to-amplitude modulation (AM/AM) and amplitude-to-phase modulation (AM/PM) distortion  \cite{N. N. Moghadam 2018}. The notations $x_{b,n}$ and $z_{b,n}$ represent  the $n$-th element of the beamformed signal $\mathbf{x}_{b} \triangleq [x_{b,1}, x_{b,2}, \ldots, x_{b,N_{\mathrm{t}}}]^{T}\in \mathbb{C}^{N_{\mathrm{t}}}$ and the nonlinear distorted signal $\mathbf{z}_{b} \triangleq [z_{b,1}, z_{b,2}, \ldots, z_{b,N_{\mathrm{t}}}]^{T} \in \mathbb{C}^{N_{\mathrm{t}}}$ at the $b$-th BS, respectively.

In order to simplify the analysis of the nonlinear distortion's influence on system performance and facilitate the development of a beamforming design algorithm, we employ Bussgang's theorem in \cite{E. Bjornson 2021} to decompose the nonlinear output signal of PAs into a scaled and rotated linear term with an additive distortion term, which can be expressed as
\begin{equation}
\label{eq:distortion_model}
\mathbf{z}_{b}=\mathbf{G}_{b}\mathbf{x}_{b}+\mathbf{d}_{b}, ~\forall b,
\end{equation}
where $\mathbf{G}_{b}\in \mathbb{C}^{N_{\mathrm{t}}\times N_{\mathrm{t}}}$ is the Bussgang gain matrix representing the linear gain applied to the input signal and the vector $\mathbf{d}_{b}\triangleq[d_{b,1}, d_{b,2}, \ldots, d_{b,N_{\mathrm{t}}}]^{T}\in \mathbb{C}^{N_{\mathrm{t}}}$ is the nonlinear distortion component over all $N_{\mathrm{t}}$ antenna elements. The Bussgang gain matrix $\mathbf{G}_{b}$ is a key parameter in this decomposition, which is defined as $\mathbf{G}_{b}\triangleq\mathbf{C}_{z_{b},x_{b}}\mathbf{C}_{x_{b},x_{b}}^{-1}$ with $\mathbf{C}_{z_{b},x_{b}}\triangleq\mathbb{E}\{\mathbf{z}_{b}\mathbf{x}_{b}^{H}\}$ representing the cross-correlation matrix between the nonlinear output $\mathbf{z}_{b}$ and the input $\mathbf{x}_{b}$, and $\mathbf{C}_{x_{b},x_{b}}\triangleq\mathbb{E}\{\mathbf{x}_{b}\mathbf{x}_{b}^{H}\}$ representing the correlation matrix of the input signal. Through further derivations, the explicit expression for the linear gain matrix $\mathbf{G}_{b}$ can be provided as
\begin{equation}
\label{G_b}
\mathbf{G}_{b}\triangleq\mathbf{C}_{z_{b},x_{b}}\mathbf{C}_{x_{b},x_{b}}^{-1}=\beta_{1}\mathbf{I}_{N_{\mathrm{t}}}+2\beta_{3}\mathrm{diag}\{\mathbf{W}_{b}\mathbf{W}_{b}^{H}\},~\forall b,
\end{equation}
which is a diagonal matrix that depends on the beamforming matrix $\mathbf{W}_{b}$.
By applying Bussgang's theorem, the distortion term $\mathbf{d}_{b}$ is introduced to capture the residual nonlinear term that cannot be explained by the linear term $\mathbf{G}_{b}\mathbf{x}_{b}$. As a result,  $\mathbf{d}_{b}$  is uncorrelated with the input signal, i.e. $\mathbb{E}\{\mathbf{d}_{b}\mathbf{x}_{b}^{H}\}=\mathbf{0}$.
Since the nonlinear distortion term $\mathbf{d}_{b}$ is a zero-mean random variable, its covariance matrix can be derived as
\begin{equation}
\label{Cd_b}
\mathbf{C}_{\mathrm{d},b}\triangleq\mathbb{E}\{\mathbf{d}_{b}\mathbf{d}_{b}^{H}\}=2\beta_{3}(\mathbf{W}_{b}\mathbf{W}_{b}^{H}\odot|\mathbf{W}_{b}\mathbf{W}_{b}^{H}|^{2})\beta_{3}^{\ast},~\forall b,
\end{equation}
which involves the beamforming matrix $\mathbf{W}_b$.
To further simplify the analysis, it is assumed, without loss of generality, that the distortion components generated by different BS are mutually uncorrelated in the cell-free network.

After linear beamforming, nonlinear amplification, and propagation through the wireless channel, the received signal at the $k$-th UE can be written as
\begin{equation}
\begin{aligned}
\label{eq:received signal distortion1}
y_{k} &= \sum_{b=1}^{B}\mathbf{h}_{b,k}^{H}\mathbf{z}_{b} + n_{k}\\
&= \sum_{b=1}^{B}\mathbf{h}_{b,k}^{H}(\mathbf{G}_{b}\mathbf{x}_{b}+\mathbf{d}_{b})+n_{k}, ~\forall k,
\end{aligned}
\end{equation}
where $\mathbf{h}_{b,k}\in \mathbb{C}^{N_{\mathrm{t}}}$ denotes the channel from the $b$-BS to the $k$-th UE, and $n_{k}\sim\mathcal{CN}(0,\sigma^{2}_{k})$ is the additive white Gaussian noise (AWGN) at the $k$-th UE. The received signal can be further described as a combination of useful signal, multiuser interference, nonlinear distortion, and noise, i.e.,
\begin{equation}
\begin{aligned}
\label{eq:received signal distortion2}
y_{k}=&\sum_{b=1}^{B}\mathbf{h}_{b,k}^{H}\mathbf{G}_{b}\mathbf{w}_{b,k}s_{k} + \sum_{b=1}^{B}\sum_{j\neq k}^{K}\mathbf{h}_{b,k}^{H}\mathbf{G}_{b}\mathbf{w}_{b,j}s_{j}\\
&\hspace{1cm}+\sum_{b=1}^{B}\mathbf{h}_{b,k}^{H}\mathbf{d}_{b} + n_{k}, ~\forall k.
\end{aligned}
\end{equation}
Thus, based on the aforementioned assumption that the distortion terms of BSs are uncorrelated, the signal-to-interference-noise-and-distortion ratio (SINDR) of the $k$-th UE can be calculated as
\begin{equation}
\label{eq:SINR_concise}
\gamma_{k}\!=\!\!\frac{|\!\sum_{b=1}^{B}\!\mathbf{h}_{b,k}^{H}\mathbf{G}_{b}\mathbf{w}_{b,k}|^{2}}{\sum_{j\neq k}^{K}\!|\!\sum_{b=1}^{B}\!\mathbf{h}_{b,k}^{H}\mathbf{G}_{b}\mathbf{w}_{b,j}|^{2} \!\!+\!\!\sum_{b=1}^{B}\!\mathbf{h}_{b,k}^{H}\mathbf{C}_{\mathrm{d},b}\mathbf{h}_{b,k}\!+\!\sigma_{k}^{2}},~\forall k.
\end{equation}

The SINDR expression above reveals that the presence of the linear amplification gains $\{\mathbf{G}_{b}\}_{b=1}^{B}$ and nonlinear distortion covariance matrices $\{\mathbf{C}_{\mathrm{d},b}\}_{b=1}^{B}$ induced by imperfect PAs will significantly impact the SINDR performance.
By analyzing the expressions for $\{\mathbf{G}_{b}\}_{b=1}^{B}$ and $\{\mathbf{C}_{\mathrm{d},b}\}_{b=1}^{B}$ in \eqref{G_b} and \eqref{Cd_b}, it is evident that both of them are strongly dependent on the beamforming matrices $\{\mathbf{W}_{b}\}_{b=1}^{B}$.
This highlights the importance of designing appropriate beamforming matrices as a pre-processing measure to mitigate the nonlinear distortion impact of practical PAs on the cell-free network, which is the primary focus of this work

\subsection{Problem Formulation and Transformation}
In order to mitigate the nonlinear distortion effect, we aim to jointly design the beamforming matrices $\{\mathbf{W}_{b}\}_{b=1}^{B}$ of all $B$ BSs to maximize the achievable sum-rate under the transmit power constraint of each BS. The optimization problem can be mathematically formulated as
\begin{subequations}\label{eq:Problem_formulation}
\begin{align}
\label{eq:Problem_formulation_a}
\max_{\{\mathbf{W}_{b}\}_{b=1}^{B}}\!\!\!\!&~~~\sum_{k=1}^{K}\log_{2}(1+\gamma_{k})\\
\!\!\!\!\!\!\!\!\mathrm{s.t.}
\label{eq:Problem_formulation_b}
&~~~~\|\mathbf{W}_{b}\|_{F}^{2}\leq P_{\mathrm{t}},~\forall b,
\end{align}
\end{subequations}
where $P_{\mathrm{t}}$ is the power budget of each BS.
Obviously, the complicated objective function \eqref{eq:Problem_formulation_a} with $\log(\cdot)$ and fractional terms greatly hinders the algorithm development. Thus, before solving this problem, we apply the fractional programming (FP) method, along with some equivalent transformations, to convert the intricate objective function into a more tractable polynomial expression, which is derived as follows.

Firstly, by employing the Lagrangian dual reformulation and introducing auxiliary variable $\bm{\mu}=[\mu_{1}, \mu_{2}, \ldots, \mu_{K}]^{T}$ \cite{FP}, \cite{H. Li 2020 Hybrid}, the objective function \eqref{eq:Problem_formulation_a} can be transformed to
\begin{equation}
\label{eq:sum log}
\begin{aligned}
&\sum_{k=1}^{K}\log_{2}(1\!+\!\mu_{k})-\!\!\sum_{k=1}^{K}\mu_{k}\\
&+\sum_{k=1}^{K}\frac{(1\!+\!\mu_{k})|\sum_{b=1}^{B}\mathbf{h}_{b,k}^{H}\mathbf{G}_{b}\mathbf{w}_{b,k}|^{2}}{\sum_{j=1}^{K}\!|\sum_{b=1}^{B}\mathbf{h}_{b,k}^{H}\mathbf{G}_{b}\mathbf{w}_{b,j}|^{2} \!+\!\sum_{b=1}^{B}\!\mathbf{h}_{b,k}^{H}\mathbf{C}_{\mathrm{d},b}\mathbf{h}_{b,k}\!\!+\!\sigma_{k}^{2}},\\
\end{aligned}
\end{equation}
which is equivalent to the objective function \eqref{eq:Problem_formulation_a} when the auxiliary variable $\mu_{k}$ has the optimal value as
\begin{equation}
\label{eq:mu}
\mu_{k}^{\star}\!=\!\!\frac{|\!\sum_{b=1}^{B}\!\mathbf{h}_{b,k}^{H}\mathbf{G}_{b}\mathbf{w}_{b,k}|^{2}}{\sum_{j\neq k}^{K}\!|\!\sum_{b=1}^{B}\!\mathbf{h}_{b,k}^{H}\mathbf{G}_{b}\mathbf{w}_{b,j}|^{2} \!\!+\!\!\sum_{b=1}^{B}\!\mathbf{h}_{b,k}^{H}\mathbf{C}_{\mathrm{d},b}\mathbf{h}_{b,k}\!+\!\sigma_{k}^{2}},\forall k.
\end{equation}
However, the sum of fractional terms in \eqref{eq:sum log} still hinders a straightforward solution. Next, to facilitate the following optimization for variables $\{\mathbf{W}_{b}\}_{b=1}^{B}$, we further apply the quadratic transform to convert the third term in \eqref{eq:sum log} into
\begin{equation} \label{eq:quadratic transform}
2\sqrt{1+\mu_{k}}\Re\Big\{\zeta_{k}^{\ast}\sum_{b=1}^{B}\mathbf{h}_{b,k}^{H}\mathbf{G}_{b}\mathbf{w}_{b,k}\Big\}-|\zeta_{k}|^{2}D_{k}, ~\forall k,
\end{equation}
where for notation simplicity we define
\begin{equation}
D_{k}\triangleq \sum_{j=1}^{K}|\sum_{b=1}^{B}\mathbf{h}_{b,k}^{H}\mathbf{G}_{b}\mathbf{w}_{b,j}|^{2} +\sum_{b=1}^{B}\mathbf{h}_{b,k}^{H}\mathbf{C}_{\mathrm{d},b}\mathbf{h}_{b,k}+\sigma_{k}^{2}, ~\forall k,
\end{equation}
and $\zeta_{k}$ is the $k$-th element of the auxiliary variable $\bm{\zeta}=[\zeta_{1}, \zeta_{2}, \ldots, \zeta_{K}]^{T}$. The expression \eqref{eq:quadratic transform} is equivalent to the last fractional term in \eqref{eq:sum log} when $\zeta_{k}$ has following optimal value
\begin{equation}
\label{eq:t}
\zeta_{k}^{\star}=\frac{\sqrt{1+\mu_{k}}\sum_{b=1}^{B}\mathbf{h}_{b,k}^{H}\mathbf{G}_{b}\mathbf{w}_{b,k}}{D_{k}}, ~\forall k.
\end{equation}

Based on the above formula derivation, the objective function \eqref{eq:Problem_formulation_a} can be reformulated as
\begin{equation} \label{eq: reformulated objective}
\begin{aligned}
&{\sum_{k=1}^{K}}\Big(\log_{2}(1+\mu_{k})-\mu_{k}\\
&\quad+ \big(2\sqrt{1+\mu_{k}}\Re\big\{\zeta_{k}^{\ast}\sum_{b=1}^{B}\mathbf{h}_{b,k}^{H}\mathbf{G}_{b}\mathbf{w}_{b,k}\big\}-|\zeta_{k}|^{2}D_{k}\big)\Big).
\end{aligned}
\end{equation}
In order to facilitate the subsequent problem optimization, the expression (\ref{eq: reformulated objective}) can be rewritten as the following concise form
\begin{equation} \label{eq:concise form_objective}
\sum_{k=1}^{K}\big(\log_{2}(1+\mu_{k})-\mu_{k}-|\zeta_{k}|^{2}\sigma_{k}^{2}\big)+\delta,
\end{equation}
where we define
\begin{equation}
\label{eq:delta}
\begin{aligned}
\delta\triangleq & \sum_{k=1}^{K}\Big(2\sqrt{1+\mu_{k}}\Re\Big\{\zeta_{k}^{\ast}\sum_{b=1}^{B}\mathbf{h}_{b,k}^{H}\mathbf{G}_{b}\mathbf{w}_{b,k}\Big\}\\
& \hspace{0.1cm}-|\zeta_{k}|^{2}\sum_{j=1}^{K}\big|\sum_{b=1}^{B}\mathbf{h}_{b,k}^{H}\mathbf{G}_{b}\mathbf{w}_{b,j}\big|^{2}\!\!-\!|\zeta_{k}|^{2}\sum_{b=1}^{B}\mathbf{h}_{b,k}^{H}\mathbf{C}_{\mathrm{d},b}\mathbf{h}_{b,k}\Big).
\end{aligned}
\end{equation}
Roughly speaking, $\delta$ can be deemed as a summation of the transformed SINDRs of $K$ UEs, where the ratio form of each SINDR is converted to the difference between useful power and multiuser interference along with nonlinear distortion. Up until now, the original objective function \eqref{eq:Problem_formulation_a} has been converted into a more solvable form. Following this, we will delve into the algorithms that can further resolve this problem, aiming to determine the most effective approach.

\subsection{Motivation of Distributed Beamforming Design}
In the conventional centralized beamforming design, we can employ the iterative optimization of the auxiliary variables $\bm{\mu}, \bm{\zeta}$ and beamforming matrices $\{\mathbf{W}_{b}\}_{b=1}^{B}$. Observing the above equivalent objective function derived in \eqref{eq:concise form_objective}, it is evident that with $\bm{\mu}$ and $\bm{\zeta}$ fixed, the objective function for optimizing $\{\mathbf{W}_{b}\}_{b=1}^{B}$ can be simplified to $\delta$ given in \eqref{eq:delta}, i.e., including only its relevant terms. Therefore, we can recast the optimization problem as
\begin{subequations}\label{eq:Problem_formulation_delta}
\begin{align}
\label{eq:Problem_formulation_delta_a}
\max_{\{\mathbf{W}_{b}\}_{b=1}^{B}}\!\!\!\!&~~~ \delta\\
\!\!\!\!\!\!\!\!\mathrm{s.t.}
&~~~\|\mathbf{W}_{b}\|_{F}^{2}\leq P_{\mathrm{t}}, ~\forall b.
\end{align}
\end{subequations}
It can be seen that within the surrogate objective function $\delta$, the complicated nonlinear distortion parameter matrices $\{\mathbf{G}_{b}\}_{b=1}^{B}$ and $\{\mathbf{C}_{\mathrm{d},b}\}_{b=1}^{B}$ presented in \eqref{G_b} and \eqref{Cd_b} are also functions of the beamforming matrices $\{\mathbf{W}_{b}\}_{b=1}^{B}$, making it very difficult to solve \eqref{eq:Problem_formulation_delta}. A promising solution is to introduce auxiliary variables to decouple the high-order terms involving $\{\mathbf{W}_{b}\}_{b=1}^{B}$ in the objective function \eqref{eq:Problem_formulation_delta_a}, transforming them into quadratic and linear terms with respect to $\{\mathbf{W}_{b}\}_{b=1}^{B}$ and the auxiliary variables. The MM algorithm can then be applied to reformulate the penalty term into a convex form, thereby enabling an iterative optimization of several sub-problems until convergence.

However, due to a substantial number of UEs and BSs in cell-free systems and the complicated nonlinear distortion model pertinent to beamforming matrices, this centralized approach encounters critical bottlenecks: Extremely large information exchange overhead and ultra-high computational complexity. Specifically, the process of BSs uploading all channel state information (CSI) and the central processor distributing all the final beamforming matrix results leads to a large overhead for transferring $2N_{\mathrm{t}}KB$ double-precision numbers. This centralized beamforming design algorithm requires a heavy computational load for calculating an $N_{\mathrm{t}}KB$-dimensional beamforming vector and an $N_{\mathrm{t}}KB\times N_{\mathrm{t}}KB$-dimensional auxiliary matrix.
To address these difficulties, distributed signal processing and optimization have emerged as promising techniques for this challenging distortion-aware beamforming design.
Thanks to the distributed architecture and local computational resources at each BS, cell-free networks are ideally suited for efficiently handling distributed beamforming design tasks.

\begin{figure*}[!t]
  \centering
  \vspace{-0.0 cm}
  \subfigure[Ring topology.]{\includegraphics[width= 2.8 in]{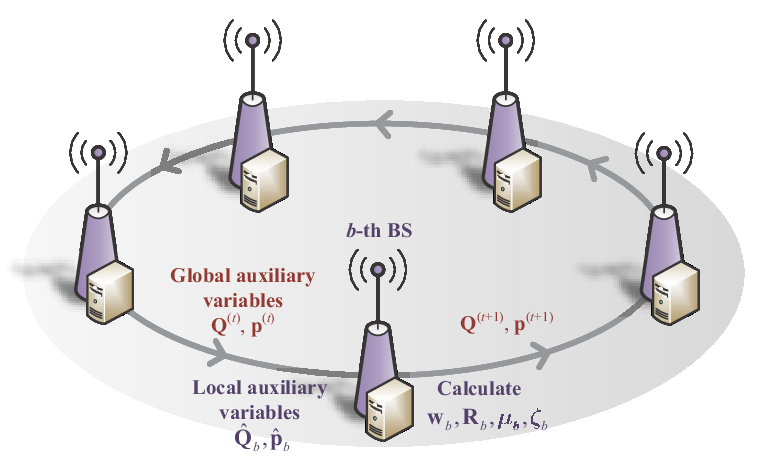}}
  \hspace{0.6cm}
  \subfigure[Star topology.]{\includegraphics[width= 2.8 in]{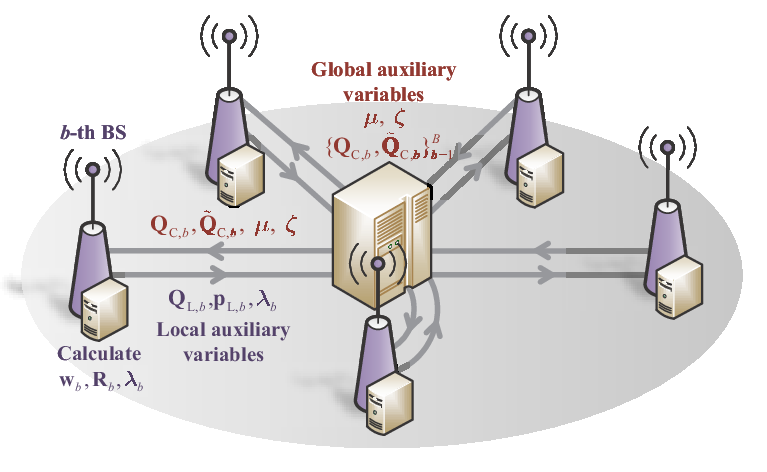}}
  \caption{Illustration of distributed beamforming designs based on the proposed topologies.}
  \label{fig:decentralized}
  \vspace{-0.2 cm}
\end{figure*}

The principle of the distributed beamforming design essentially aims to allow BSs to locally perform the iterative optimization of the auxiliary variables $\bm{\mu}, \bm{\zeta}$ and beamforming matrices $\{\mathbf{W}_{b}\}_{b=1}^{B}$ in a distributed manner.
Within the distributed beamforming design schemes, instead of uploading/downloading high-dimensional channel information/beamforming matrices and conducting high-complexity centralized design, each BS only needs to collect its own CSI, exchange a small amount of global information, and execute light-weight calculations. This approach is more scalable, robust, and efficient.
Therefore, it is imperative to develop appropriate distributed beamforming design algorithms for cell-free networks to achieve a better trade-off among network capacity, information exchange overhead, and computational complexity.
To achieve this goal, we will introduce two distributed beamforming designs based on ring topology and star topology, respectively, as illustrated in Fig. \ref{fig:decentralized}.
In particular, when we are in the pursuit of easing the computational load on the central processor, the ring-topology-based fully-distributed scheme can be employed; conversely, if we aim to utilize the superior computation capability of the central processor for achieving better sum-rate performance, the star-topology-based partially-distributed scheme can be adopted.

\section{Ring-Topology-based Fully-Distributed Beamforming Design Algorithm}
In this section, we develop a fully-distributed beamforming design algorithm based on a ring topology without the central processor, as shown in Fig. \ref{fig:decentralized} (a). To be specific, after receiving the information of global auxiliary variables from the previous BS on the ring, the current BS calculates its own optimal beamforming matrix, updates the global auxiliary variables, and then relays them to the next neighboring BS. Each BS successively designs its conditionally optimal beamforming with updated global auxiliary variables, with one iteration corresponding to the complete processing by each BS, continuing until convergence is achieved. In the following, we first derive global auxiliary variables to enable efficient information sharing. Then regarding the local beamforming design at each BS, we employ a penalty-MM-based algorithm to handle the high-order terms in the objective function and alternately solve the sub-problems.

\subsection{Information Sharing}
With sufficient global information, each BS can design its own beamforming by alternatively optimizing variables $\mathbf{W}_{b}$, $\bm{\mu}_{b}$, and $\bm{\zeta}_{b}$. When $ \bm{\mu}_{b}$ and $ \bm{\zeta}_{b}$ are fixed, the original objective function can also be simplified to $\delta$, and the local beamforming design can be formulated as
\begin{subequations}\label{eq:Problem_formulation_deltainformationsharing}
\begin{align}
\label{eq:Problem_formulation_deltainformationsharing_a}
\max_{\mathbf{W}_{b}}&~~~~~~ \delta\\
\mathrm{s.t.}
&~~~\|\mathbf{W}_{b}\|_{F}^{2}\leq P_{\mathrm{t}}.
\end{align}
\end{subequations}
However, the expression of the objective function $\delta$ shown in \eqref{eq:delta} includes information of CSI and amplification distortion, which is challenging to obtain from the other BSs, rendering it imperative to share partial information between the BSs. Instead of exchanging high-dimensional CSI, beamforming matrices and distortion matrices, we attempt to directly pack and share the low-dimensional information of the useful, interference and nonlinear distortion beams related to all the BSs to achieve efficient information sharing. This information is irrelevant to the number of transmit antennas, thereby greatly reducing the overhead cost of the backhaul link.

\newcounter{TempEqCnt1}
\setcounter{TempEqCnt1}{\value{equation}}
\setcounter{equation}{18}
\begin{figure*}[b]
\hrulefill
\begin{equation}\label{eq:delta_Qplong}
\begin{aligned}
   \delta = &\sum_{k=1}^{K}\Big(2\sqrt{1+\mu_{k}}\Re\big\{\zeta_{k}^{\ast}(\overbrace{\underbrace{\sum_{l\neq b}^{B}\mathbf{h}_{l,k}^{H}\mathbf{G}_{l}\mathbf{w}_{l,k}}_{\textrm{useful signal of}\atop\textrm{the other BSs}}}^{\mathbf{\widehat{Q}}_{b}(k,k)}   +\underbrace{\mathbf{h}_{b,k}^{H}\mathbf{G}_{b}\mathbf{w}_{b,k}}_{\textrm{useful signal of}\atop\textrm{the $b$-th BSs}})\big\}
 -|\zeta_{k}|^{2}\sum_{j=1}^{K}\big|(\overbrace{\underbrace{\sum_{l\neq b}^{B}\mathbf{h}_{l,k}^{H}\mathbf{G}_{l}\mathbf{w}_{l,j}}_{\textrm{interference of}\atop\textrm{the other BSs}}}^{\mathbf{\widehat{Q}}_{b}(k,j)}+\underbrace{\mathbf{h}_{b,k}^{H}\mathbf{G}_{b}\mathbf{w}_{b,j}}_{\textrm{interference of}\atop\textrm{the $b$-th BSs}})\big|^{2} \\ &\hspace{1cm}-|\zeta_{k}|^{2}(\overbrace{\underbrace{\sum_{l\neq b}^{B}\mathbf{h}_{l,k}^{H}\mathbf{C}_{\mathrm{d},l}\mathbf{h}_{l,k}}_{\textrm{distortion of}\atop\textrm{the other BSs}}}^{\mathbf{\widehat{p}}_{b}(k)}+\underbrace{\mathbf{h}_{b,k}^{H}\mathbf{C}_{\mathrm{d},b}\mathbf{h}_{b,k}}_{\textrm{distortion of}\atop \textrm{the  $b$-th BSs}}\Big).
 \end{aligned}
\end{equation}
\setcounter{equation}{21}
\begin{equation}\label{eq:delta_Qplong_matrix}
  \delta=  \sum_{k=1}^{K}\Big(2\sqrt{1+\mu_{k}}\Re\big\{\zeta_{k}^{\ast}(\mathbf{\widehat{Q}}_{b}(k,k)+\mathbf{h}_{b,k}^{H}\mathbf{G}_{b}\mathbf{w}_{b,k})\big\}-|\zeta_{k}|^{2}\sum_{j=1}^{K}\big|\mathbf{\widehat{Q}}_{b}(k,j)+\mathbf{h}_{b,k}^{H}\mathbf{G}_{b}\mathbf{w}_{b,j}\big|^{2} -|\zeta_{k}|^{2}\big(\mathbf{\widehat{p}}_{b}(k)+\mathbf{h}_{b,k}^{H}\mathbf{C}_{\mathrm{d},b}\mathbf{h}_{b,k}\big)\Big).
\end{equation}
\end{figure*}

To be specific, from the perspective of the $b$-th BS, we can rewrite the objective function, i.e. $\delta$, into the equivalent form as shown in \eqref{eq:delta_Qplong} at the bottom of the this page, in which each summation term is partitioned into two components, i.e., the information of the $b$-th BS and the other BSs, respectively.
Within this objective function, the $b$-th BS can obtain the local information by estimating its own CSI $\{\mathbf{h}_{b,k}\}_{k=1}^{K}$, while the information of the other BSs is acquired from the previous BS by global information sharing mechanism.
Based on the above derivation, we can define local auxiliary variables $\mathbf{\widehat{Q}}_{b}$ and $\mathbf{\widehat{p}}_{b}$ containing the information of the other BSs, excluding that of the $b$-th BS, whose $(k,j)$-th element and $k$-th element are respectively calculated by
\setcounter{equation}{19}
\begin{eqnarray}
\label{eq:Q_hat}
\vspace{-0.0cm}
\mathbf{\widehat{Q}}_{b}(k,j)&=&\sum\nolimits_{l\neq b}^{B}\mathbf{h}_{l,k}^{H}\mathbf{G}_{l}\mathbf{w}_{l,j},~\forall k, j,\\
\label{eq:p_hat}\mathbf{\widehat{p}}_{b}(k)&=&\sum\nolimits_{l\neq b}^{B}\mathbf{h}_{l,k}^{H}\mathbf{C}_{\mathrm{d},l}\mathbf{h}_{l,k},~\forall k.
\end{eqnarray}
Particularly, $\mathbf{\widehat{Q}}_{b}$ has the useful signal information on its diagonal, the interference signal on the off-diagonal elements, and $\mathbf{\widehat{p}}_{b}$ contains the distortion for the UEs. Based on this definition, the objective function can be further rewritten as \eqref{eq:delta_Qplong_matrix} illustrated at the bottom of this page. Thus, after acquiring the other BSs' information $\mathbf{\widehat{Q}}_{b}$ and $\mathbf{\widehat{p}}_{b}$ and having its own information $\{\mathbf{h}_{b,k}\}_{k=1}^{K}$, the $b$-th BS can execute its local beamforming design.

\newcounter{TempEqCnt2}
\setcounter{TempEqCnt2}{\value{equation}}
\begin{figure*}[b]
\vspace{-0.3cm}
\hrulefill
\setcounter{equation}{26}
\begin{equation}
\label{eq:delta_hat_b}
\begin{aligned}
\widehat{\delta}_{b}=&\sum_{k=1}^{K}\!\!\Big(2\sqrt{1\!+\!\mu_{k}}\Re\{\zeta_{k}^{\ast}\mathbf{h}_{b,k}^{H}\mathbf{G}_{b}\mathbf{w}_{b,k}\}
\!-\!|\zeta_{k}|^{2}\mathbf{h}_{b,k}^{H}\mathbf{C}_{\mathrm{d},b}\mathbf{h}_{b,k}\!-\!|\zeta_{k}|^{2}\!\sum_{j=1}^{K}\!\!\big(2\Re\{\mathbf{\widehat{Q}}_{b}(k,j)^{\ast}\mathbf{h}_{b,k}^{H}\mathbf{G}_{b}\mathbf{w}_{b,j}\}\!+\!\mathbf{h}_{b,k}^{H}\mathbf{G}_{b}\mathbf{w}_{b,j}\mathbf{w}_{b,j} ^{H}\mathbf{G}_{b}^{H}\mathbf{h}_{b,k}\big)\Big).\\
\end{aligned}
\end{equation}
\end{figure*}

While the $b$-th BS only needs auxiliary variables $\mathbf{\widehat{Q}}_{b}$ and $\mathbf{\widehat{p}}_{b}$ of other BSs for calculating its own beamforming, the information sharing mechanism around the ring should include the information of all BSs, since the information of the $b$-th BS is also necessary for other BSs.
Therefore, in order to facilitate more efficient information sharing, we introduce a global auxiliary matrix $\mathbf{Q}\in \mathbb{C}^{K\times K}$ and a global auxiliary vector $\mathbf{p}\in \mathbb{C}^{K}$, both of which aggregate the information from all BSs. The global auxiliary matrix $\mathbf{Q}$ contains the receiving useful signal and interference serving for each UE, whose $(k,j)$-th element can be calculated as\setcounter{equation}{22}
\vspace{-0.1cm}
\begin{equation}
\label{eq:Q}
\mathbf{Q}(k,j)=\sum\nolimits_{l=1}^{B}\mathbf{h}_{l,k}^{H}\mathbf{G}_{l}\mathbf{w}_{l,j}, ~\forall k, j.
\end{equation}
Similarly, the global auxiliary vector $\mathbf{p}\in \mathbb{C}^{K}$ represents the distortion, with its $k$-th element derived as
\begin{equation}
\label{eq:p}
\mathbf{p}(k)=\sum\nolimits_{l=1}^{B}\mathbf{h}_{l,k}^{H}\mathbf{C}_{\mathrm{d},l}\mathbf{h}_{l,k}, ~\forall k.
 \vspace{-0.1cm}
\end{equation}


For the $b$-th BS, after receiving the global auxiliary matrix $\mathbf{{Q}}$ and vector $\mathbf{{p}}$ from the previous BS, it first subtracts the components associated with itself to obtain the local auxiliary matrix   $\mathbf{\widehat{Q}}_{b}$ and vector $\mathbf{\widehat{p}}_{b}$, whose $(k,j)$-th element and $k$-th elements are respectively calculated as
\vspace{-0.1cm}
\begin{eqnarray}
\label{eq:Q_hat}
  \mathbf{\widehat{Q}}_{b}(k,j)
  &=&\mathbf{Q}(k,j)-\mathbf{h}_{b,k}^{H}\mathbf{G}_{b}\mathbf{w}_{b,j}, ~\forall k, j, \\
\label{eq:p_hat}
  \mathbf{\widehat{p}}_{b}(k)
  &=&\mathbf{p}(k)-\mathbf{h}_{b,k}^{H}\mathbf{C}_{\mathrm{d},b}\mathbf{h}_{b,k}, ~\forall k,
\end{eqnarray}
where $\mathbf{h}_{b,k}^{H}\mathbf{G}_{b}\mathbf{w}_{b,j}$ and $\mathbf{h}_{b,k}^{H}\mathbf{C}_{\mathrm{d},b}\mathbf{h}_{b,k}$ are the computation outcomes of the previous round of iteration at the $b$-th BS, temporarily kept at the local BS.

After simplification and retaining only the terms related to the beamforming matrix $\mathbf{W}_{b}$, we can obtain a new objective function for the beamforming design of the $b$-th BS as $\widehat{\delta}_{b}$ shown in (\ref{eq:delta_hat_b}) at the bottom of this page. \setcounter{equation}{27}
Hence, the local optimization problem at the $b$-th BS is formulated as
\begin{subequations}
\label{eq:Problem_formulation_local_initial}
\begin{align}
\label{eq:Problem_formulation_local_initial_a}
\max_{\mathbf{W}_{b}}\!&\ \ \  \widehat{\delta}_{b}\\
\mathrm{s.t.}
\label{eq:Problem_formulation_local_initial_b}
&\ \ \|\mathbf{W}_{b}\|_{F}^{2}\leq P_{\mathrm{t}},
\end{align}
\end{subequations}
where the $b$-th BS is solely responsible for optimizing its own beamforming matrix, mitigating its nonlinear distortion effect.

After solving the problem \eqref{eq:Problem_formulation_local_initial} and having the optimized beamforming $\{\mathbf{w}_{b,k}\}_{k=1}^{K}$, the $b$-th BS will update the global auxiliary variables
\begin{eqnarray}
\label{eq:Q_update}
  \mathbf{Q}(k,j) &=& \mathbf{\widehat{Q}}_{b}(k,j) + \mathbf{h}_{b,k}^{H}\mathbf{G}_{b}\mathbf{w}_{b,j}, ~\forall  k, j,\\
\label{eq:p_update}
  \mathbf{p}(k) &=& \mathbf{\widehat{p}}_{b}(k) + \mathbf{h}_{b,k}^{H}\mathbf{C}_{\mathrm{d},b}\mathbf{h}_{b,k}, ~\forall k,
\end{eqnarray}
and share them with the next BS for the subsequent iteration.

In summary, after receiving the global auxiliary
variables $\mathbf{Q},\mathbf{p}$ from the preceding BS, the current BS calculates local auxiliary variables $\mathbf{\widehat{Q}}_{b}$ and  $\mathbf{\widehat{p}}_{b}$, optimizes local beamforming, updates global auxiliary
variables and transfers them to the next BS.
Each BS sequentially executes the above procedure until the result converges.
Having described the mechanism of information sharing and distributed computation for the ring-topology-based cell-free network, next we turn to focus on the local beamforming design for the $b$-th BS.

\subsection{Local Beamforming Design}
At the $b$-th BS, after receiving the global variables $\mathbf{Q}$ and $\mathbf{p}$ from the preceding BS, its current task is to resolve problem \eqref{eq:Problem_formulation_local_initial}.
It can be  noticed from \eqref{eq:delta_hat_b} that the objective function $\widehat{\delta}_{b}$ still remains intractable,
because it is a mixture of high-order and low-order terms with respect to variable $\mathbf{W}_{b}$ and the distortion parameter matrices $\mathbf{G}_{b}$ and $\mathbf{C}_{\mathrm{d},b}$ involve complex non-convex operations with variable $\mathbf{W}_{b}$, such as Hadamard product and $|\cdot|$. Hence, we propose to introduce auxiliary variables to tackle these difficulties.

First, we convert the beamforming matrix to be optimized into a vector form as
\begin{equation}
\mathbf{w}_{b} = \mathrm{vec}(\mathbf{W}_{b}).
\end{equation}
Then, the linear amplification gain matrix $\mathbf{G}_{b}$ can be transformed into an equivalent form as
\begin{equation}
\label{eq:G}
\mathbf{G}_{b}=\beta_{1}\mathbf{I}_{N_{\mathrm{t}}}+2\beta_{3}\mathbf{E}_{1}^{T}(
\mathbf{w}_{b}\mathbf{w}_{b}^{H}\odot\mathbf{I}_{N_{\mathrm{t}}K})\mathbf{E}_{1},
\end{equation}
where $\mathbf{E}_{1}\triangleq\mathbf{1}_{K}\otimes \mathbf{I}_{N_{\mathrm{t}}}$ is the constant matrix. Similarly, the covariance matrix of the nonlinear distortion term, i.e.,  $\mathbf{C}_{\mathrm{d},b}$, can be equivalently expressed as
\vspace{-0.1cm}
\begin{equation}
\label{eq:Cd}
\mathbf{C}_{\mathrm{d},b}=2\beta_{3}(\mathbf{F}_{b}\odot|\mathbf{F}_{b}|^{2})\beta_{3}^{\ast},
\end{equation}
where for brevity we define
\begin{equation}
\label{eq:F}
\mathbf{F}_{b}\triangleq\mathbf{E}_{1}^{T}(
\mathbf{w}_{b}\mathbf{w}_{b}^{H}\odot(\mathbf{I}_{K}\otimes\mathbf{1}_{N_{\mathrm{t}}\times N_{\mathrm{t}}}))\mathbf{E}_{1}.
\end{equation}
To facilitate the algorithm development, we further convert the objective function \eqref{eq:delta_hat_b} into the following concise form
\begin{equation}
\begin{aligned}
\label{eq:delta_hat_b_concise}
\widehat{\delta}_{b}\triangleq &\ 2\Re\{\mathbf{h}_{b}^{H}\mathbf{E}_{2}\mathbf{\overline{G}}_{b}\mathbf{w}_{b}\}
-2\Re\{\mathbf{e}_{b}^{H}\mathbf{\overline{G}}_{b}\mathbf{w}_{b}\}\\
&\hspace{0.5cm}-\mathbf{w}_{b}^{H}\mathbf{\overline{G}}_{b}^{H}\mathbf{E}_{3,b}\mathbf{\overline{G}}_{b}\mathbf{w}_{b}-
\mathbf{h}_{b}^{H}\mathbf{E}_{4}^{H}\mathbf{\overline{C}}_{\mathrm{d},b}\mathbf{E}_{4}\mathbf{h}_{b},\\
\end{aligned}
\end{equation}
where for brevity we define the following matrices and vectors
\begin{equation}
\begin{aligned}\!\!\mathbf{H}_{b}&\triangleq[\mathbf{h}_{b,1}, \mathbf{h}_{b,2}, \ldots, \mathbf{h}_{b,K}]\in \mathbb{C}^{N_{\mathrm{t}}\times K}, \mathbf{h}_{b} \triangleq \text{vec}(\mathbf{H}_{b}),\\
\!\!\mathbf{E}_{2}&\triangleq(\mathbf{I}_{N_{\mathrm{t}}K}+\mathrm{diag}\{\bm{\mu}\}\otimes\mathbf{I}_{N_{\mathrm{t}}})^{\frac{1}{2}}(\mathrm{diag}\{\bm{\zeta}^{\ast}\}\otimes\mathbf{I}_{N_{\mathrm{t}}}),\\
\!\!\mathbf{E}_{3,b}&\triangleq\mathbf{I}_{K}\otimes\mathbf{H}_{b}\mathrm{diag}\{\bm{\zeta}\}\mathrm{diag}\{\bm{\zeta}^{*}\}\mathbf{H}_{b}^{H},\\
\!\!\mathbf{E}_{4}&\triangleq\mathrm{diag}\{\bm{\zeta}\}\otimes\mathbf{I}_{N_{\mathrm{t}}},~~\mathbf{e}_{b,j}\triangleq\sum_{k=1}^{K}|\zeta_{k}|^{2}\mathbf{\widehat{Q}}_{b}(k,j)\mathbf{h}_{b,k},\\
\!\!\mathbf{e}_{b}&\triangleq[\mathbf{e}_{b,1}^{T}, \mathbf{e}_{b,j}^{T}, \ldots, \mathbf{e}_{b,K}^{T}]^{T}\in \mathbb{C}^{N_{\mathrm{t}}K},\\
\!\!\mathbf{\overline{G}}_{b}&\triangleq\mathbf{I}_{K}\otimes\mathbf{G}_{b},\quad
\mathbf{\overline{C}}_{\mathrm{d},b}\triangleq \mathbf{I}_{K}\otimes\mathbf{C}_{\mathrm{d},b}.
\end{aligned}
\end{equation}
Considering the non-convexity of \eqref{eq:G} and \eqref{eq:Cd} regarding $\mathbf{w}_{b}$, we propose to introduce another auxiliary variable $\mathbf{R}_{b}=\mathbf{w}_{b}\mathbf{w}_{b}^H$ to decouple the high-order terms, and rewrite the matrices $\mathbf{G}_{b}$ and $\mathbf{C}_{\mathrm{d},b}$ as
\begin{eqnarray}
\label{eq:G2}
\mathbf{G}_{b} &=& \beta_{1}\mathbf{I}_{N_{\mathrm{t}}}+2\beta_{3}\mathbf{E}_{1}^{T}(\mathbf{R}_{b}\odot\mathbf{I}_{N_{\mathrm{t}}K})\mathbf{E}_{1},\\
\label{eq:Cd2}
\mathbf{C}_{\mathrm{d},b} &=& 2\beta_{3}(\mathbf{\overline{F}}_{b}\odot|\mathbf{\overline{F}}_{b}|^{2})\beta_{3}^{\ast},
\end{eqnarray}
where $\mathbf{\overline{F}}_{b}=\mathbf{E}_{1}^{T}(\mathbf{R}_{b}\odot(\mathbf{I}_{K}\otimes\mathbf{1}_{N_{\mathrm{t}}\times N_{\mathrm{t}}}))\mathbf{E}_{1}$ is a linear function of $\mathbf{R}_{b}$. Thus, the beamforming design problem at the $b$-th BS can be reformulated as
\begin{subequations}
\label{eq:Problem_formulation_R}
\begin{align}
\label{eq:Problem_formulation_R_a}
\max_{\mathbf{w}_{b},\mathbf{R}_{b}}\!&\ \ \widehat{\delta}_{b}\\
\mathrm{s.t.}
\label{eq:Problem_formulation_R_b}
&\ \ \|\mathbf{w}_{b}\|_{2}^{2}\leq P_{\mathrm{t}}, \\
\label{eq:Problem_formulation_R_c}
&\ \ \mathbf{R}_{b}=\mathbf{w}_{b}\mathbf{w}_{b}^H.
\end{align}
\end{subequations}
Then, the equality constraint \eqref{eq:Problem_formulation_R_c} is relaxed and added to the objective function as a penalty term, thus the problem \eqref{eq:Problem_formulation_R} is converted to the following penalized problem
\begin{subequations}
\label{eq:Problem_formulation_R_penalty}
\begin{align}
\label{eq:Problem_formulation_R_penalty_a}
\min_{\mathbf{w}_{b},\mathbf{R}_{b}}\!\!&\ \ -\widehat{\delta}_{b}+\rho\|\mathbf{R}_{b}-\mathbf{w}_{b}\mathbf{w}_{b}^H\|_{F}^{2}\\
\label{eq:Problem_formulation_R_penalty_b}
\!\!\!\!\!\!\mathrm{s.t.}
&\ \ \|\mathbf{w}_{b}\|_{2}^{2}\leq P_{\mathrm{t}},
\end{align}
\end{subequations}
where $\rho>0$ represents the penalty coefficient and $\mathbf{R}_{b}$ is an $N_\text{t}K\times N_\text{t}K$-dimensional matrix.

To solve this more tractable bi-variate problem, the current BS alternately optimizes its own variables $\mathbf{w}_{b}$ and $\mathbf{R}_{b}$, calculates $\bm{\mu}_{b}$ and $\bm{\zeta}_{b}$, updates the global auxiliary variables $\mathbf{Q}$ and $\mathbf{p}$, and then transfers them to the next adjacent BS for the next iteration, which can reduce local computational complexity.
The detailed algorithm derivation is described as follows.

\subsubsection{Optimize $\mathbf{w}_{b}$}
The objective function $-\widehat{\delta}_{b}$ is convex with respect to $\mathbf{w}_{b}$ according to its definition in \eqref{eq:delta_hat_b_concise}, but the penalty term is non-convex. To address this issue, we employ the power constraint \eqref{eq:Problem_formulation_R_penalty_b} to construct an upper bound of the penalty term as
\begin{equation}\label{eq:use Pt for penalty}
\|\mathbf{R}_{b}-\mathbf{w}_{b}\mathbf{w}_{b}^H\|_{F}^{2}\leq\mathrm{Tr}(\mathbf{R}_{b}^{H}\mathbf{R}_{b})-2\mathbf{w}_{b}^H\mathbf{R}_{b}\mathbf{w}_{b}+P_{\mathrm{t}}\mathbf{w}_{b}^H\mathbf{w}_{b}.
\end{equation}
In \eqref{eq:use Pt for penalty}, the second term $-2\mathbf{w}_{b}^H\mathbf{R}_{b}\mathbf{w}_{b}$ is non-convex with respect to $\mathbf{w}_{b}$. We propose to seek for a linear surrogate function for this concave quadratic term utilizing the MM algorithm \cite{MM}, \cite{R. Liu}. Specifically, by employing a first-order Taylor expansion, a linear surrogate function for $-2\mathbf{w}_{b}^H\mathbf{R}_{b}\mathbf{w}_{b}$ at the current point $\mathbf{w}_{b}^{(t)}$ is given by
\begin{equation}\label{eq:linear surr}
\!\!\!-2\mathbf{w}_{b}^H\mathbf{R}_{b}\mathbf{w}_{b} \leq2(\mathbf{w}_{b}^{(t)})^{H}\mathbf{R}_{b}\mathbf{w}_{b}^{(t)}-4\Re\{(\mathbf{w}_{b}^{(t)})^{H}\mathbf{R}_{b}\mathbf{w}_{b}\},
\end{equation}
where $\mathbf{w}_{b}^{(t)}=\text{vec}(\mathbf{W}_{b}^{(t)})$ is the vector form of the beamforming matrix at the current point.
Using the results in \eqref{eq:use Pt for penalty}, \eqref{eq:linear surr} and ignoring irrelevant terms, the sub-problem for updating $\mathbf{w}_{b}$ can be expressed as the following convex form
\begin{subequations}\label{eq:Problem_formulation_opt_W}
\begin{align}
\label{eq:Problem_formulation_opt_W_a}
\!\!\!\!\!\min_{\mathbf{w}_{b}}&~\mathbf{w}_{b}^{H}(\mathbf{\overline{G}}_{b}^{H}\mathbf{E}_{3,b}\mathbf{\overline{G}}_{b}+\rho P_{\mathrm{t}}\mathbf{I}_{N_\text{t}K})\mathbf{w}_{b}\nonumber\\
\!\!\!\!&\quad+2\Re\big\{(\mathbf{e}_{b}^{H}\mathbf{\overline{G}}_{b}-\mathbf{h}_{b}^{H}\mathbf{E}_{2}\mathbf{\overline{G}}_{b}-2\rho(\mathbf{w}_{b}^{(t)})^{H}\mathbf{R}_{b})\mathbf{w}_{b}\big\} \!\!\!\!\!\!\\
\!\!\!\!\mathrm{s.t.}
\label{eq:Problem_formulation_opt_W_b}
&\ \|\mathbf{w}_{b}\|_{2}^{2}\leq P_{\mathrm{t}},
\end{align}
\end{subequations}
which can be easily tackled using either off-the-shelf solver CVX or using the solution derived in the following.

The Lagrangian function of the problem \eqref{eq:Problem_formulation_opt_W} is given by
\begin{equation}
\begin{aligned}
\mathcal{L}_{\mathrm{w},b}(\mathbf{w}_{b},\eta_{b})&=\mathbf{w}_{b}^{H}(\mathbf{\overline{G}}_{b}^{H}\mathbf{E}_{3,b}\mathbf{\overline{G}}_{b}+\rho P_{\mathrm{t}}\mathbf{I}_{N_\text{t}K})\mathbf{w}_{b}\\
& +2\Re\big\{(\mathbf{e}_{b}^{H}\mathbf{\overline{G}}_{b}-\mathbf{h}_{b}^{H}\mathbf{E}_{2}\mathbf{\overline{G}}_{b}-2\rho(\mathbf{w}_{b}^{(t)})^{H}\mathbf{R}_{b})\mathbf{w}_{b}\big\}
\\&+\eta_{b}(\|\mathbf{w}_{b}\|_{2}^{2}-P_{\mathrm{t}}),
\end{aligned}
\end{equation}
where $\eta_{b}\geq 0$ is the Lagrangian multiplier.
Then, it can be reformulated into the following concise form
\begin{equation}
\mathcal{L}_{\mathrm{w},b}(\mathbf{w}_{b},\eta_{b})=\mathbf{w}_{b}^{H}\mathbf{C}_{\mathrm{w},b}\mathbf{w}_{b}+2\Re\{\mathbf{c}_{\mathrm{w},b}^{H}\mathbf{w}_{b}\}-\eta_{b} P_{\mathrm{t}},
\end{equation}
where for brevity we define
\begin{equation}
\begin{aligned}
\mathbf{C}_{\mathrm{w},b}\triangleq&\ \mathbf{\overline{G}}_{b}^{H}\mathbf{E}_{3,b}\mathbf{\overline{G}}_{b}+\rho P_{\mathrm{t}}\mathbf{I}_{N_\text{t}K}+\eta_{b}\mathbf{I}_{N_\text{t}K},\\
\mathbf{c}_{\mathrm{w},b}\triangleq&\ (\mathbf{e}_{b}^{H}\mathbf{\overline{G}}_{b}-\mathbf{h}_{b}^{H}\mathbf{E}_{2}\mathbf{\overline{G}}_{b}-2\rho(\mathbf{w}_{b}^{(t)})^{H}\mathbf{R}_{b})^{H}.
\end{aligned}
\end{equation}
By setting the partial derivative of $\mathcal{L}_{\mathrm{w},b}(\mathbf{w}_{b},\eta_{b})$ with respect to $\mathbf{w}_{b}$ to zero, i.e., $\frac{\partial\mathcal{L}_{\mathrm{w},b}(\mathbf{w}_{b},\eta_{b})}{\partial \mathbf{w}_{b}}=0$, the optimal beamforming vector can be obtained as follows
\begin{equation}
\label{closed_form_w}
\mathbf{w}_{b}^{\star}=-\mathbf{C}_{\mathrm{w},b}^{-1}\mathbf{c}_{\mathrm{w},b}.
\end{equation}
Next, in order to determine the Lagrangian multiplier $\eta_{b}$, let $\mathbf{w}_{b}(\eta_{b})$ denote the right-hand side of \eqref{closed_form_w}. Two conditions must be considered: if $\|\mathbf{w}_{b}(0)\|_{2}^{2}\leq P_{\mathrm{t}}$, then  $\mathbf{w}_{b}^{\star}=\mathbf{w}_{b}(0)$; otherwise,  it follows that $\|\mathbf{w}_{b}^{\star}(\eta_{b})\|_{2}^{2}= P_{\mathrm{t}}$, with $\eta_{b}$ being computed through one-dimensional search techniques such as the bisection method.

\subsubsection{Optimize $\mathbf{R}_{b}$}
Given the beamforming vector $\mathbf{w}_{b}$, the sub-problem of solving $\mathbf{R}_{b}$ can be formulated as
\begin{equation}
\label{eq:Problem_formulation_opt_R_penalty}
    \min_{\mathbf{R}_{b}}\ \ -\widehat{\delta}_{b}+\rho\|\mathbf{R}_{b}-\mathbf{w}_{b}\mathbf{w}_{b}^H\|_{F}^{2}.
\end{equation}
Recall that $\widehat{\delta}_{b}$ in \eqref{eq:delta_hat_b_concise} is a quadratic function of $\mathbf{G}_{b}$ which is a linear function of $\mathbf{R}_{b}$ shown in \eqref{eq:G2}, while the matrix $\mathbf{C}_{\mathrm{d},b}$ is a complex cubic function of $\mathbf{R}_{b}$ as in \eqref{eq:Cd2}. In order to tackle this intricate term, we propose to decouple the Hadamard product terms in \eqref{eq:Cd2} and alternately update them. In particular, with the obtained current solution of $\mathbf{R}_{b}^{(t)}$, we first update the term
\begin{equation}
\label{eq:transform Fb} |\mathbf{\overline{F}}_{b}^{(t+1)}|^2=|\mathbf{E}_{1}^{T}(\mathbf{R}_{b}^{(t)}\odot(\mathbf{I}_{K}\otimes\mathbf{1}_{N_{\mathrm{t}}\times N_{\mathrm{t}}}))\mathbf{E}_{1}|^2,
\end{equation}
and then reformulate the matrix $\mathbf{C}_{\mathrm{d},b}$ as
\begin{equation}
\label{eq:transform Cd}
   \mathbf{C}_{\mathrm{d},b}= 2\beta_3\big[\big(\mathbf{E}_{1}^{T}(\mathbf{R}_{b}\odot(\mathbf{I}_{K}\otimes\mathbf{1}_{N_{\mathrm{t}}\times N_{\mathrm{t}}}))\mathbf{E}_{1}\big)\odot|\mathbf{\overline{F}}_{b}^{(t+1)}|^{2}\big]\beta_{3}^{\ast},
\end{equation}
which is also a linear function regarding $\mathbf{R}_{b}$.
Now, the optimization problem for $\mathbf{R}_{b}$ can be formulated as a convex problem as
\begin{equation}\label{eq:Problem_formulation_opt_R}
\begin{aligned}
\!\!\! \min_{\mathbf{R}_{b}}&\ \mathbf{w}_{b}^{H}\mathbf{\overline{G}}_{b}^{H}\mathbf{E}_{3,b}\mathbf{\overline{G}}_{b}\mathbf{w}_{b}\!+\!
\Re\{(2\mathbf{e}_{b}^{H}\mathbf{\overline{G}}_{b}-\mathbf{h}_{b}^{H}\mathbf{E}_{2}\mathbf{\overline{G}}_{b})\mathbf{w}_{b}\}\\
&\ +\mathbf{h}_{b}^{H}\mathbf{E}_{4}^{H}\mathbf{\overline{C}}_{\mathrm{d},b}\mathbf{E}_{4}\mathbf{h}_{b}+
\rho\|\mathbf{R}_{b}-\mathbf{w}_{b}\mathbf{w}_{b}^H\|_{F}^{2}.
\end{aligned}
\end{equation}
The problem can be efficiently solved either directly using the CVX solver or by computing the derivative with respective to $\mathbf{R}_{b}$ to derive a closed-form solution as
\begin{equation}
\begin{aligned}
\mathrm{vec}(\mathbf{R}_{b})=-((\mathbf{C}_{\mathrm{R},b}+\rho\mathbf{I}_{N_{\mathrm{t}}^{2}K^{2}})^{-1}\mathbf{c}_{\mathrm{R},b})^{\ast},\\
\end{aligned}
\end{equation}
where the detailed expressions for $\mathbf{C}_{\mathrm{R},b}$ and $\mathbf{c}_{\mathrm{R},b}$ are shown in \eqref{eq:CRcR} with the proof provided in Appendix A.
\newcounter{TempEqCnt}
\setcounter{TempEqCnt}{\value{equation}}
\setcounter{equation}{52}
\begin{figure*}[b]
\hrulefill
\begin{equation}\label{eq:miu_Q}
\begin{aligned}
   \mu_{b,k}^{\star}=\frac{|\mathbf{\widehat{Q}}_{b}(k,k)+\mathbf{h}_{b,k}^{H}\mathbf{G}_{b}\mathbf{w}_{b,k}|^{2}}{\sum_{j\neq k}^{K}|\mathbf{\widehat{Q}}_{b}(k,j)+\mathbf{h}_{b,k}^{H}\mathbf{G}_{b}\mathbf{w}_{b,j}|^{2} +\mathbf{\widehat{p}}_{b}(k)+\mathbf{h}_{b,k}^{H}\mathbf{C}_{\mathrm{d},b}\mathbf{h}_{b,k}+\sigma_{k}^{2}}, ~\forall k,
\end{aligned}
\end{equation}
\begin{equation}\label{eq:zeta_Q}
\zeta_{b,k}^{\star}=\frac{\sqrt{1+\mu_{b,k}}(\mathbf{\widehat{Q}}_{b}(k,k)+\mathbf{h}_{b,k}^{H}\mathbf{G}_{b}\mathbf{w}_{b,k})}
{\sum_{j=1}^{K}|\mathbf{\widehat{Q}}_{b}(k,j)+\mathbf{h}_{b,k}^{H}\mathbf{G}_{b}\mathbf{w}_{b,j}|^{2} +\mathbf{\widehat{p}}_{b}(k)+\mathbf{h}_{b,k}^{H}\mathbf{C}_{\mathrm{d},b}\mathbf{h}_{b,k}+\sigma_{k}^{2}}, ~\forall k.
\end{equation}
\end{figure*}

After calculating the variables $\mathbf{w}_{b}$, $\mathbf{R}_{b}$ and utilizing the known information $\mathbf{\widehat{Q}}_{b}, \mathbf{\widehat{p}}_{b}$, the auxiliary variables $\bm{\mu}_{b}=[\mu_{b,1}, \mu_{b,2}, \ldots, \mu_{b,K}]^{T}$ and $\bm{\zeta}_{b}=[\zeta_{b,1}, \zeta_{b,2}, \ldots, \zeta_{b,K}]^{T}$ are updated as \eqref{eq:miu_Q} and \eqref{eq:zeta_Q} shown at the bottom of the next page.

\begin{algorithm}[!t]
\caption{The ring-topology-based fully-distributed distortion-aware beamforming design algorithm.}
    \begin{algorithmic}[1]
    \begin{small}
    \REQUIRE $\{\mathbf{H}_{b}\}_{b=1}^{B}$ for the corresponding BSs, $T$, $B$.
    \ENSURE $\{\mathbf{w}_{b}^{\star}\}_{b=1}^{B}$.
            \STATE {Initialize $\{\mathbf{w}_{b},\mathbf{R}_{b}\}_{b=1}^{B}$ for local BSs.}
            \STATE {Initialize $\mathbf{Q}$ and $\mathbf{p}$ for information relaying.}
            \STATE {Set $t=1$.}
            \WHILE {$t\leq T$ and no convergence of the objective function}
                   \STATE {\textbf{At the $b$-th BS, $b$:=$(t\ \mathrm{mod}\ B)+1$}:}
                   \STATE {Receive $\mathbf{Q}^{(t)}$, $\mathbf{p}^{(t)}$ from the previous BS.}
                   \STATE {Calculate $\mathbf{\widehat{Q}}_{b}^{(t)},\mathbf{\widehat{p}}_{b}^{(t)}$ by \eqref{eq:Q_hat}, \eqref{eq:p_hat}.}
                   \STATE {Obtain beamforming  $\mathbf{w}_{b}^{(t+1)}$ by solving \eqref{eq:Problem_formulation_opt_W}.}
                   \STATE{Update $|\mathbf{F}_{b}^{(t+1)}|^2$ by \eqref{eq:transform Fb}.}
                   \STATE {Update  $\mathbf{R}_{b}^{(t+1)}$ by solving \eqref{eq:Problem_formulation_opt_R}.}
                   \STATE {Update $\bm{\mu}_{b}^{(t+1)}$ by \eqref{eq:miu_Q}.}
                   \STATE {Update $\bm{\zeta}_{b}^{(t+1)}$ by \eqref{eq:zeta_Q}.}
                   \STATE {Update   $\mathbf{Q}^{(t+1)},\mathbf{p}^{(t+1)}$ by \eqref{eq:Q_update} and \eqref{eq:p_update}. }
                   \STATE {Relay  $\mathbf{Q}^{(t+1)},\mathbf{p}^{(t+1)}$ to the $(b+1)$-th BS.}
                   \STATE {$t:=t+1$.}
            \ENDWHILE
            \STATE {Return $\{\mathbf{w}_{b}^{\star}\}_{b=1}^{B}$.}
    \end{small}
    \vspace{-0.0 cm}
    \end{algorithmic}
    \vspace{-0.0 cm}
\end{algorithm}

\subsection{Summary, Complexity Analysis, and Backhaul Signaling Overhead Analysis}
\subsubsection{Summary}
Based on the above development, the proposed ring-topology-based distributed beamforming algorithm to solve the problem \eqref{eq:Problem_formulation} is straightforward and summarized in Algorithm 1. First, the local variables $\{\mathbf{w}_{b},\mathbf{R}_{b}\}_{b=1}^{B}$ and global auxiliary variables $\mathbf{Q}, \mathbf{p}$ are appropriately initialized. Then, after receiving the global variables $\mathbf{Q}$ and $\mathbf{p}$ from the previous BS and calculating the local auxiliary variables $\mathbf{\widehat{Q}}_{b},\mathbf{\widehat{p}}_{b}$, the current BS will update the beamforming vector $\mathbf{w}_{b}$ and the auxiliary variables $\mathbf{R}_{b}$, $\bm{\mu}_{b}$,  $\bm{\zeta}_{b}$. Eventually, the current BS updates its relevant information in the global variables and relays them to the next BS. In essence, all the BSs take turns iterating once until convergence, forming a ring topology. Regarding convergence, since solving each sub-problem optimally results in a non-increasing objective function and a lower bound on the objective function value exists, it can be concluded that the algorithm will converge to a local optimum. A detailed convergence analysis can be found at https://rangliu0706.github.io/.


\subsubsection{Complexity Analysis}
In addition, we briefly analyze the computational complexity of the proposed ring-topology-based distributed beamforming algorithm.
Specifically, the computational complexity predominantly lies in calculating beamforming vector $\mathbf{w}_{b}$ by \eqref{eq:Problem_formulation_opt_W} and auxiliary matrix $\mathbf{R}_{b}$ by \eqref{eq:Problem_formulation_opt_R}. At each BS, concerning the convex problem \eqref{eq:Problem_formulation_opt_W} with an $N_{\mathrm{t}}K\times 1$-dimensional variable $\mathbf{w}_{b}$ to be optimized and an $N_{\mathrm{t}}K\times 1$-dimensional second-order cone (SOC) constraint, the computational complexity is of order $\mathcal{O}\{2\sqrt{2}N_{\mathrm{t}}^{3}K^{3}+\sqrt{2}N_{\mathrm{t}}K\}$. Similarly, the computational complexity of optimizing problem \eqref{eq:Problem_formulation_opt_R} with an $N_{\mathrm{t}}K\times N_{\mathrm{t}}K$-dimensional variable $\mathbf{R}_{b}$ is of order $\mathcal{O}\{N_{\mathrm{t}}^{4}K^{4}\}$. Therefore, the total computational complexity of Algorithm 1 is of order $\mathcal{O}\{N_{\mathrm{iter}}(N_{\mathrm{t}}^{4}K^{4}+2\sqrt{2}N_{\mathrm{t}}^{3}K^{3}+\sqrt{2}N_{\mathrm{t}}K)\}$, where $N_{\mathrm{iter}}$ is the number of iterations.

\subsubsection{Backhaul Signaling Overhead Analysis}
This ring-topology-based fully-distributed algorithm significantly reduces backhaul signaling overhead, as it avoids the need to exchange all high-dimensional CSI among BSs. Instead, the algorithm transfers only low-dimensional global variables $\mathbf{Q}$ and $\mathbf{p}$, which encapsulate the information related to the useful signal, interference, and distortion. Quantitatively speaking, the total required backhaul signaling overhead of the proposed scheme is $N_{\mathrm{iter}}(K^{2}+K)$ double-precision numbers.

\section{Star-Topology-based Partially-Distributed Beamforming Design Algorithm}
In this section, we focus on a partially-distributed scheme based on the star topology, as illustrated in Fig. \ref{fig:decentralized} (b). Unlike the ring topology where the beamforming design does not involve a central processor, this scheme leverages the computational capability of the central processor to enhance performance, albeit at the cost of increased backhaul signaling overhead and higher computational complexity.
Specifically, after BSs download global auxiliary variables from the central processor, they design their beamforming matrices in parallel and then upload their local computation results back to the central processor for further information aggregation. This process is iteratively repeated in coordination between the BSs and the central processor until convergence is achieved. In the following subsections, we first introduce global and local auxiliary variables to facilitate the download and upload of information. Then, a consensus-ADMM-based algorithm is employed to establish consensus between the local BSs and the central processor, while utilizing the same penalty-MM-based algorithm described in Section III-B to alleviate the nonlinear distortion effect. 

\vspace{-0.3 cm}
\subsection{Information Sharing and Consensus}
In the star-topology-based system, the information sharing is performed between the local BSs and the central processor.
Similar to the discussion in Section III-A, in
order to realize efficient information sharing, we still define low-dimensional
local auxiliary variables $\mathbf{Q}_{\mathrm{L},b}\in \mathbb{C}^{K\times K}$ and $\mathbf{p}_{\mathrm{L},b}\in \mathbb{C}^{K}$, $b=1,\ldots, B$. These auxiliary variables are individually computed at each local BS as follows
\begin{eqnarray}
\label{eq:QLb}
   \mathbf{Q}_{\mathrm{L},b} &=&\mathbf{H}_{b}^{H}\mathbf{G}_{b}\mathbf{W}_{b},~\forall b,\\
\label{eq:pLb}
   \mathbf{p}_{\mathrm{L},b} &=&\mathrm{vec}(\mathrm{diag}\{\mathbf{H}_{b}^{H}\mathbf{C}_{\mathrm{d},b}\mathbf{H}_{b}\}),~\forall b.
\end{eqnarray}
Similarly, $\mathbf{Q}_{\mathrm{L},b}\in \mathbb{C}^{K\times K}$  captures the information of both useful signal and interference, while $\mathbf{p}_{\mathrm{L},b}\in \mathbb{C}^{K}$  contains the distortion information.
Subsequently, the BSs upload the local information to the central processor for aggregation and to update auxiliary variables $\bm{\mu}$ and $\bm{\zeta}$. The central processor, in turn, is responsible for distributing the aggregated information to the  BSs for distributed beamforming designs. To facilitate efficient information aggregation and distribution, we introduce a set of global auxiliary variables $\mathbf{Q}_{\mathrm{C},b} \in \mathbb{C}^{K\times K} $, $b=1,\ldots, B$, associated with each local BS.
The purpose of aggregating the local variables $\{\mathbf{Q}_{\mathrm{L},b}\}_{b=1}^{B}$  and distributing the global variables $\{\mathbf{Q}_{\mathrm{C},b}\}_{b=1}^{B}$ is to facilitate the exchange of interference information among the BSs, which are coupled in the beamforming design. In contrast, as the distortion components are independent across BSs, only the local variables $\{\mathbf{p}_{\mathrm{L},b}\}_{b=1}^{B} $ need to be uploaded to the central processor for updating $\bm{\mu}$ and $\bm{\zeta}$, without the need to define corresponding global variables for their exchange across BSs.

Although $\{\mathbf{Q}_{\mathrm{C},b}\}^{B}_{b=1}$ are computed at the central processor and $\{\mathbf{Q}_{\mathrm{L},b}\}^{B}_{b=1}$ are calculated at local BSs, both contain the same underlying information. Thus, it is essential to align the local and central results to achieve consensus, which is the key to obtaining a globally optimal solution. To enforce this, we impose consistency constraints at both the center and local, i.e.,$\mathbf{Q}_{\mathrm{L},b}=\mathbf{Q}_{\mathrm{C},b}$, which can be addressed through a consensus-based ADMM framework.

\subsection{Global Information Aggregation and Local Beamforming Design}
\subsubsection{Global Information Aggregation}
After the central processor receives $\{\mathbf{Q}_{\mathrm{L},b}\}^{B}_{b=1}$ from all local BSs, the information aggregation problem, which involves computing the global auxiliary variables $\{\mathbf{Q}_{\mathrm{C},b}\}_{b=1}^{B}$, can be formulated as
\begin{subequations}\label{eq:Problem_formulation_Q_global}
\begin{align}
\label{eq:Problem_formulation_Q_global_a}
\!\!\!\!\!\!\min_{\{\mathbf{Q}_{\mathrm{C},b}\}_{b=1}^{B}}&-\delta_{\mathrm{c}}\\
\label{eq:Problem_formulation_Q_global_b}
\mathrm{s.t.}
\ \ \ &\mathbf{Q}_{\mathrm{C},b}=\mathbf{Q}_{\mathrm{L},b}, ~\forall b,
\end{align}
\end{subequations}
where the objective function $\delta_{\mathrm{c}}$ is expressed as
\begin{equation}
\begin{aligned}
\label{eq:delta_global}
\delta_{\mathrm{c}}&=\sum_{k=1}^{K}\Big(2\sqrt{1+\mu_{k}}\Re\big\{\zeta_{k}^{\ast}\sum\nolimits_{b=1}^{B}\mathbf{Q}_{\mathrm{C},b}(k,k)\big\}\\
&\hspace{1.4cm}-|\zeta_{k}|^{2}\sum\nolimits_{j=1}^{K}\big|\sum\nolimits_{b=1}^{B}\mathbf{Q}_{\mathrm{C},b}(k,j)\big|^{2}\Big),
\end{aligned}
\end{equation}
and \eqref{eq:Problem_formulation_Q_global_b} is the global-local consensus constraint.
We first reformulate the equality constraints in \eqref{eq:Problem_formulation_Q_global_b} in vector form as
\begin{equation}
\label{eq:q_tranform}
\mathbf{q}_{\mathrm{C},b}=\mathbf{q}_{\mathrm{L},b}, ~\forall b,
\end{equation}
where $\mathbf{q}_{\mathrm{C},b}=\mathrm{vec}(\mathbf{Q}_{\mathrm{C},b})$,  $\mathbf{q}_{\mathrm{L},b}=\mathrm{vec}(\mathbf{Q}_{\mathrm{L},b})$. Then, the information aggregation problem can be solved by optimizing the problem with its augmented Lagrangian (AL) function
\begin{equation}\label{eq:Problem_formulation_Q_global_AL}
\begin{aligned}
\hspace{-0.28cm}\min_{\{\mathbf{Q}_{\mathrm{C},b}\}_{b=1}^{B}}&-\delta_{\mathrm{c}}
+\frac{\varrho}{2}\sum_{b=1}^{B}\|\mathbf{q}_{\mathrm{C},b}-\mathbf{q}_{\mathrm{L},b}+\frac{\bm{\lambda}_{b}}{\varrho}\|_{2}^{2},
\end{aligned}
\end{equation}
where $\bm{\lambda}_{b}\in \mathbb{C}^{K\times K}$ is the dual variable.
The problem in \eqref{eq:Problem_formulation_Q_global_AL} is convex and can be efficiently solved using CVX.

The obtained $\mathbf{Q}_{\mathrm{C},b}$ contains
information of both useful signal for the $b$-th BS and the inter-BS interference to the other BSs. To facilitate the sharing of inter-BS interference information and enable the $b$-th BS to efficiently assess the interference, we also compute the aggregated interference information from all other BSs, excluding its own, as
\begin{equation}
\label{eq:tilde_Qp}
\mathbf{\widetilde{Q}}_{\mathrm{C},b}=\sum\nolimits_{l\neq b}^{B}\mathbf{Q}_{\mathrm{C},b}.
\end{equation}

Furthermore, using the computed global variables $\{\mathbf{Q}_{\mathrm{C},b}\}_{b=1}^{B}$ and the received local variables $\{\mathbf{p}_{\mathrm{L},b}\}_{b=1}^{B}$, the central processor updates $\bm{\mu}$ and $\bm{\zeta}$ as follows
\begin{eqnarray}
\label{eq:update mu}  \!\!\!\!\!\!\!\mu_{k}^{\star}\!\!\!\!\!\!&=&\!\!\!\!\!\!\frac{|\sum_{b=1}^{B}\!\mathbf{Q}_{\mathrm{C},b}(k,k)|^{2}}{\sum_{j\neq k}^{K}\!|\sum_{b=1}^{B}\!\mathbf{Q}_{\mathrm{C},b}(k,j)|^{2}\!+\!\sum_{b=1}^{B}\mathbf{p}_{\mathrm{L},b}(k)\!+\!\sigma_{k}^{2}},\forall k, \\
  \label{eq:update zeta}
  \!\!\!\!\!\!\!\zeta_{k}^{\star}\!\!\!\!\!\!&=&\!\!\!\!\!\!\frac{\sqrt{1+\mu_{k}}(\sum_{b=1}^{B}\!\mathbf{Q}_{\mathrm{C},b}(k,k))}{\sum_{j= 1}^{K}\!|\sum_{b=1}^{B}\!\mathbf{Q}_{\mathrm{C},b}(k,j)|^{2}\!+\!\sum_{b=1}^{B}\mathbf{p}_{\mathrm{L},b}(k)\!+\!\sigma_{k}^{2}},\forall k.
\end{eqnarray}
After completing the above computation, the central processor distribute the resulting $\mathbf{Q}_{\mathrm{C},b}$, $
 \widetilde{\mathbf{Q}}_{\mathrm{C},b}$, $\bm{\mu}$, and $\bm{\zeta}$ to the $b$-th local BS for further distributed beamforming design.

\subsubsection{Local Beamforming Design}
Following a similar transformation as \eqref{eq:delta_hat_b}, the objective function $\delta$ in \eqref{eq:delta} can be simplified and reformulated as
\begin{equation}
\begin{aligned}
\widetilde{\delta}_{b}=&\sum_{k=1}^{K}\Big(2\sqrt{1+\mu_{k}}\Re\{\zeta_{k}^{\ast}\mathbf{h}_{b,k}^{H}\mathbf{G}_{b}\mathbf{w}_{b,k}\}
-|\zeta_{k}|^{2}\mathbf{h}_{b,k}^{H}\mathbf{C}_{\mathrm{d},b}\mathbf{h}_{b,k}\\
&\hspace{1cm}-|\zeta_{k}|^{2}\sum\nolimits_{j=1}^{K}\big(2\Re\{\mathbf{\widetilde{Q}}_{\mathrm{C},b}(k,j)^{\ast}\mathbf{h}_{b,k}^{H}\mathbf{G}_{b}\mathbf{w}_{b,j}\}\\
&\hspace{3cm}+\mathbf{h}_{b,k}^{H}\mathbf{G}_{b}\mathbf{w}_{b,j}\mathbf{w}_{b,j} ^{H}\mathbf{G}_{b}^{H}\mathbf{h}_{b,k}\big)\Big).\\
\end{aligned}
\end{equation}
By adopting the approach in \eqref{eq:Problem_formulation_R_penalty} to manage high-order terms and introducing the consensus constraint, the local optimization problem can be reformulated as
\begin{subequations}\label{eq:Problem_formulation_Q_local}
\begin{align}
\label{eq:Problem_formulation_Q_local_a}
\min_{\mathbf{w}_{b},\mathbf{R}_{b}}\!\!&\ \ -\widetilde{\delta}_{b}
+\rho\|\mathbf{R}_{b}-\mathbf{w}_{b}\mathbf{w}_{b}^H\|_{F}^{2}\\
\label{eq:Problem_formulation_Q_local_b}
\mathrm{s.t.}
&\ \ \|\mathbf{w}_{b}\|_{2}^{2}\leq P_{\mathrm{t}},\\
\label{eq:Problem_formulation_Q_local_c}&\ \ \mathbf{H}_{b}^{H}\mathbf{G}_{b}\mathbf{W}_{b} = \mathbf{Q}_{\mathrm{C},b}.
\end{align}
\end{subequations}

Analogously, the local BSs design their own beamforming vector $\mathbf{w}_{b}$ in parallel by solving problem \eqref{eq:Problem_formulation_Q_local}.
Particularly, after transforming the consensus constraint \eqref{eq:Problem_formulation_Q_local_c} into vector form, the local optimization problem \eqref{eq:Problem_formulation_Q_local} can be reformulated with its AL function as
\begin{subequations}\label{eq:Problem_formulation_Q_local_AL}
\begin{align}
\label{eq:Problem_formulation_Q_local_AL_a}
\hspace{-0.2cm}\hspace{-0.3cm}\min_{\mathbf{w}_{b}\, \mathbf{R}_{b}}&\ -\widetilde{\delta}_{b}
+\rho\|\mathbf{R}_{b}-\mathbf{w}_{b}\mathbf{w}_{b}^H\|_{F}^{2}\nonumber\\
&\ +\frac{\varrho}{2}\|\mathbf{q}_{\mathrm{C},b}-\mathrm{vec}(\mathbf{H}_{b}^{H}\mathbf{G}_{b}\mathbf{W}_{b})+\frac{\bm{\lambda}_{b}}{\varrho}\|_{2}^{2}\\
\label{eq:Problem_formulation_Q_local_AL_b}
\mathrm{s.t.}
 &\ \ \|\mathbf{w}_{b}\|_{2}^{2}\leq P_{\mathrm{t}},
\end{align}
\end{subequations}
where $\varrho>0$ is a penalty parameter. With the fixed global variable $\mathbf{Q}_{\mathrm{C},b}$ and dual variable $\bm{\lambda}_{b}$, it can be observed that the term $\|\mathbf{q}_{\mathrm{C},b}-\mathrm{vec}(\mathbf{H}_{b}^{H}\mathbf{G}_{b}\mathbf{W}_{b})+\frac{\bm{\lambda}_{b}}{\varrho}\|_{2}^{2}$ is convex with respect to   $\mathbf{w}_{b}$ and $\mathbf{R}_{b}$, making it straightforward to handle. Thus, the optimization problem in \eqref{eq:Problem_formulation_Q_local_AL} can be solved using the same method described in Section III-B.

Besides, at the $b$-th BS, after receiving global variable $\mathbf{Q}_{\mathrm{C},b}$ and computing local variables $\mathbf{w}_{b}, \mathbf{R}_{b}$, the dual variable $\bm{\lambda}_{b}$ is updated by
\begin{equation}
\label{eq:update lamda}
\bm{\lambda}_{b}:=\bm{\lambda}_{b} + \frac{\varrho}{2}(\mathbf{q}_{\mathrm{C},b}-(\mathbf{I}_{K}\otimes\mathbf{H}_{b})^{H}\mathbf{\overline{G}}_{b}\mathbf{w}_{b}).
\end{equation}
Subsequently, with $\mathbf{w}_{b}$, $\mathbf{R}_{b}$, and $\bm{\lambda}_{b}$ determined, the local BS computes $\mathbf{Q}_{\mathrm{L},b}$ and $ \mathbf{p}_{\mathrm{L},b}$ using \eqref{eq:QLb} and \eqref{eq:pLb}, respectively. These results are then uploaded to the central processor for the  information aggregation in the next iteration.

\begin{algorithm}[!t]
\caption{The star-topology-based partially-distributed distortion-aware beamforming design algorithm.}
    \begin{algorithmic}[1]
    \begin{small}
    \REQUIRE $\{\mathbf{H}_{b}\}_{b=1}^{B}$ for the corresponding BSs, $T$, $B$.
    \ENSURE $\{\mathbf{w}_{b}^{\star}\}_{b=1}^{B}$.
            \STATE {Initialize $\{\mathbf{w}_{b},\mathbf{R}_{b},\bm{\lambda}_{b}\}_{b=1}^{B}$ for local BSs.}
            \STATE {Initialize $\{\mathbf{Q}_{\mathrm{C},b}\}_{b=1}^{B}$ for central processor.}
            \STATE {Set $t=1$.}
             \WHILE {$t\leq T$ and no convergence of the objective function}
                    \STATE {\textbf{Global: Information Aggregation}}
                    \STATE {Collect $\{\mathbf{Q}_{\mathrm{L},b}^{(t)}, \mathbf{p}_{\mathrm{L},b}^{(t)}, \bm{\lambda}_{b}^{(t)}\}_{b=1}^{B}$ from local BSs.}
                   \STATE {Update $\{\mathbf{Q}_{\mathrm{C},b}^{(t+1)}\}_{b=1}^{B}$ by solving \eqref{eq:Problem_formulation_Q_global_AL}.}
                    \STATE {Update $\{\mathbf{\widetilde{Q}}_{\mathrm{C},b}^{(t+1)}\}_{b=1}^{B}$ using \eqref{eq:tilde_Qp}.}
                   \STATE {Update $\bm{\mu}^{(t+1)}$ and $\bm{\zeta}^{(t+1)}$ by \eqref{eq:update mu} and \eqref{eq:update zeta}, respectively.}
                   \STATE {Distribute $\mathbf{Q}_{\mathrm{C},b}^{(t+1)}, \mathbf{\widetilde{Q}}_{\mathrm{C},b}^{(t+1)}, \bm{\mu}^{(t+1)}, \bm{\zeta}^{(t+1)}$ to the $b$-th BS, $\forall b$.}
                    \STATE {\textbf{Local: Parallel Beamforming Design}}
                    \FOR {$b=1, 2, \ldots, B$}
                    \STATE {Obtain $\mathbf{w}_{b}^{(t+1)}, \mathbf{R}_{b}^{(t+1)},\mathbf{F}_{b}^{(t+1)}$ by solving \eqref{eq:Problem_formulation_Q_local_AL}.}
                    \STATE {Update $\bm{\lambda}_{b}^{(t+1)}$ by \eqref{eq:update lamda}.}
                    \STATE {Calculate  $\mathbf{Q}_{\mathrm{L},b}^{(t+1)}, \mathbf{p}_{\mathrm{L},b}^{(t+1)}$ using \eqref{eq:QLb} and \eqref{eq:pLb}.}
                    \STATE {Send $\mathbf{Q}_{\mathrm{L},b}^{(t+1)}, \mathbf{p}_{\mathrm{L},b}^{(t+1)},\bm{\lambda}_{b}^{(t+1)}$ to the center.}
                   \ENDFOR
                   \STATE {$t:=t+1$.}
            \ENDWHILE
            \STATE {Return $\{\mathbf{w}_{b}^{\star}\}_{b=1}^{B}$.}
    \end{small}
    \vspace{-0.0 cm}
    \end{algorithmic}
    \vspace{-0.0 cm}
\end{algorithm}

\vspace{-0.3 cm}
\subsection{Summary, Complexity Analysis, and Backhaul Signaling Overhead Analysis}
\subsubsection{Summary}
In light of the aforementioned derivations, the proposed star-topology-based partially-distributed beamforming design algorithm to address the problem \eqref{eq:Problem_formulation} is succinctly outlined in Algorithm 2. First, the initial values of the local variables $\{\mathbf{w}_{b},\mathbf{R}_{b}, \bm{\lambda}_{b}\}_{b=1}^{B}$ and   for each BS and the global auxiliary variables $\{\mathbf{Q}_{\mathrm{C},b}\}_{b=1}^{B}$ for the central processor are properly selected. During each iteration, the central processor collects  $\{\mathbf{Q}_{\mathrm{L},b},\mathbf{p}_{\mathrm{L},b},\bm{\lambda}_{b}\}_{b=1}^{B}$ from the local BSs, updates $\{\mathbf{Q}_{\mathrm{C},b}, \mathbf{\widetilde{Q}}_{\mathrm{C},b} \}_{b=1}^{B}$, $\bm{\mu},\bm{\zeta}$, and then distributes them to corresponding BSs. With the received information, the local BSs compute their own beamforing $\mathbf{w}_{b}$ and local variables  in parallel and then send the updated information back to the center for further fusion. The central processor and local BSs alternately perform computation and information exchange until the convergence is achieved. A comprehensive analysis of the convergence can be found at https://rangliu0706.github.io/.

\subsubsection{Complexity Analysis}
In each iteration, the primary computational complexity at the central node arises from calculating the $K \times K$-dimensional variables $\{\mathbf{Q}_{\mathrm{L},b}\}_{b=1}^{B}$ in \eqref{eq:Problem_formulation_Q_global_AL}, which has a computational complexity of $\mathcal{O}\{K^{6}\}$. At the local BSs, the computational complexity mainly stems from calculating the $N_{\mathrm{t}}K$-dimensional $\mathbf{w}_{b}$ with an $N_{\mathrm{t}}K$-dimensional SOC constraint, and $N_{\mathrm{t}}K\times N_{\mathrm{t}}K$-dimensional $\mathbf{R}_{b}$ in \eqref{eq:Problem_formulation_Q_local_AL}, which is of order $\mathcal{O}\{2\sqrt{2}N_{\mathrm{t}}^{3}K^{3}+\sqrt{2}N_{\mathrm{t}}K\}$ and $\mathcal{O}\{N_{\mathrm{t}}^{4}K^{4}\}$, respectively. Consequently, the overall computational complexity of Algorithm 2 is of order $\mathcal{O}\{N_{\mathrm{iter}}(B(2\sqrt{2}N_{\mathrm{t}}^{3}K^{3}+\sqrt{2}N_{\mathrm{t}}K+N_{\mathrm{t}}^{4}K^{4})+K^{6})\}$.

\subsubsection{Backhaul Signaling Overhead Analysis}
In comparison to the ring-topology-based algorithm, the proposed star-topology-based scheme incurs a higher cost in terms of backhaul resources and computational complexity, but offers improved performance.  Concerning the backhaul signaling overhead, each iteration involves two phases of information transfer, i.e., download and upload. In the download phase, the central processor will distribute the variables $\{\mathbf{Q}_{\mathrm{C},b},\mathbf{\widetilde{Q}}_{\mathrm{C},b}\}_{b=1}^{B}$ and $\bm{\mu}, \bm{\zeta}$ to each local BS, resulting in $B(2K^{2}+2K)$ overhead. During the upload phase, every BS needs to send the results $\mathbf{Q}_{\mathrm{L},b},\mathbf{p}_{\mathrm{L},b}$ and dual variables $\bm{\lambda}_{b}$ to the center, leading to $B(2K^{2}+K)$ double-precision numbers overhead. Overall, the total backhaul signaling overhead is $N_{\mathrm{iter}}B(4K^{2}+3K)$ double-precision numbers.

\section{Simulation Results}

\begin{figure}[t]
  \centering
  \includegraphics[width= 2.7 in]{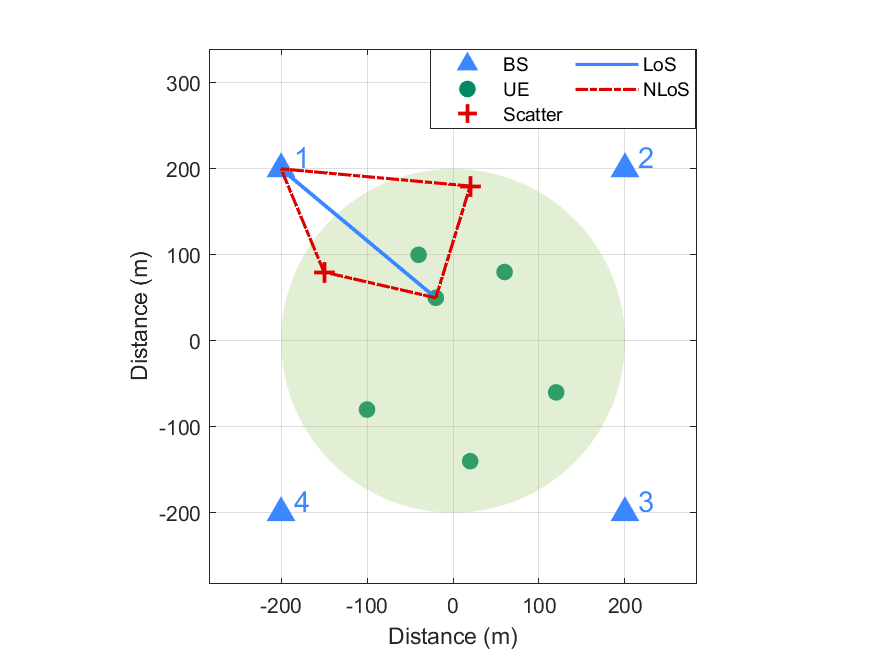}\\
  \caption{Diagram of the distribution of BSs and UEs locations.}
  \label{fig:setting}
\end{figure}

In this section, we provide simulation results to demonstrate the effectiveness of the proposed distributed distortion-aware beamforming design algorithms in alleviating the performance degradation caused by nonlinear PAs. Unless otherwise specified, as illustrated in Fig. \ref{fig:setting}, the cell-free system has $B=4$ BSs located at the coordinates $(-200\mathrm{m}, 200\mathrm{m})$, $(200\mathrm{m}, 200\mathrm{m})$, $(200\mathrm{m}, -200\mathrm{m})$, $(-200\mathrm{m}, -200\mathrm{m})$. Each BS is equipped with $N_{\mathrm{t}} = 16$ antennas, and the center carrier frequency is set to $f_{\mathrm{c}}= 28$ GHz. $K=6$ UEs randomly distributed within a circular area with a radius of $200$m.
The mmWave channel is modeled as a classic sparse multipath model. Specifically, the channel between the $b$-th BS and the $k$-th UE is given by
\begin{equation}
\begin{aligned}
\!\!\!\mathbf{h}_{b,k}=&\sum_{m=1}^{M}\alpha_{bkm}[1, e^{-j\frac{2\pi f_{\mathrm{c}}d}{c}\sin\theta_{bkm}}, \ldots,\\
 &\ \ \ \ \ \  e^{-j\frac{2\pi f_{\mathrm{c}}(N_{\mathrm{t}}-1)d}{c}\sin\theta_{bkm}}]^{T}, ~\forall b,k,
\end{aligned}
\end{equation}
where $\alpha_{bkm}=10^{-\frac{C_{0}}{10}}(\frac{r_{bkm}}{D_{0}})^{-\kappa_{m}}$ is the path loss with $r_{bkm}$ representing the propagation distance, $\kappa_{m}$ being the attenuation factor for the
$m$-th path, $C_{0}=30$ dB, $D_{0}=1$m, and $d={\lambda_{\mathrm{c}}}/{2}$ is the antenna spacing with $\lambda_{\mathrm{c}}$ known as the center carrier wavelength. We consider $M=3$ paths, including one line-of-sight (LoS) channel with the attenuation factor $\kappa_{1}=2.5$, and two non-line-of-sight (NLoS) channels with $\kappa_{2}$ and $\kappa_{3}$ randomly chosen from the range $[3, 3.5]$. Besides, for the NLoS channels, the azimuth angle is uniformly distributed within $\theta_{bkm}\in[-{\pi}/{2},{\pi}/{2}]$ and the distance is uniformly distributed within $r_{bkm}\in[200\mathrm{m}, 400\mathrm{m}]$. The parameters of the nonlinear PA model in \eqref{eq:distortion_model1} are set as $\beta_{1}=1$ and $\beta_{3}=-0.212e^{-j2.816}$ as specified in the literature \cite{E. Bjornson 2024}-\cite{M. Wu 2022}. The noise power of UEs is set as $-70$dBm.

\begin{figure}[t]
    \centering
    \subfigure[]{\includegraphics[width= 1.5 in]{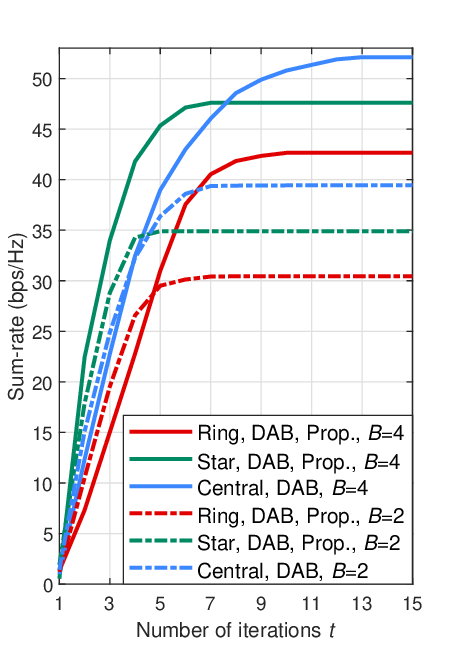}}%
    \vspace{-0.0 cm}
    \subfigure[]{\includegraphics[width= 1.5 in]{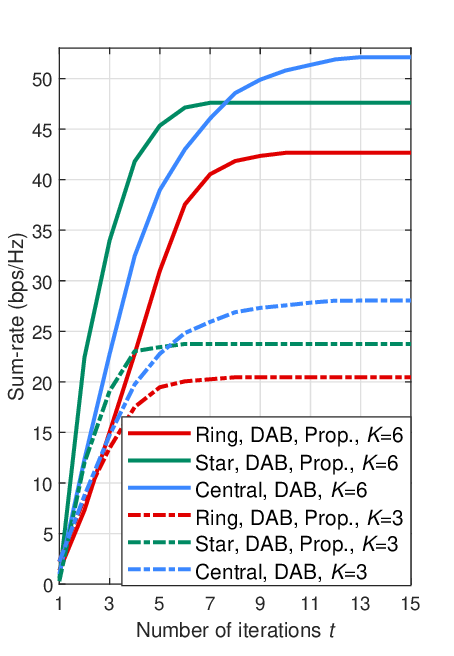}}
    \caption{Convergence performance.}
    \label{fig:convergence}
\end{figure}

\begin{figure}[t]
  \centering
  \includegraphics[width= 3.1 in]{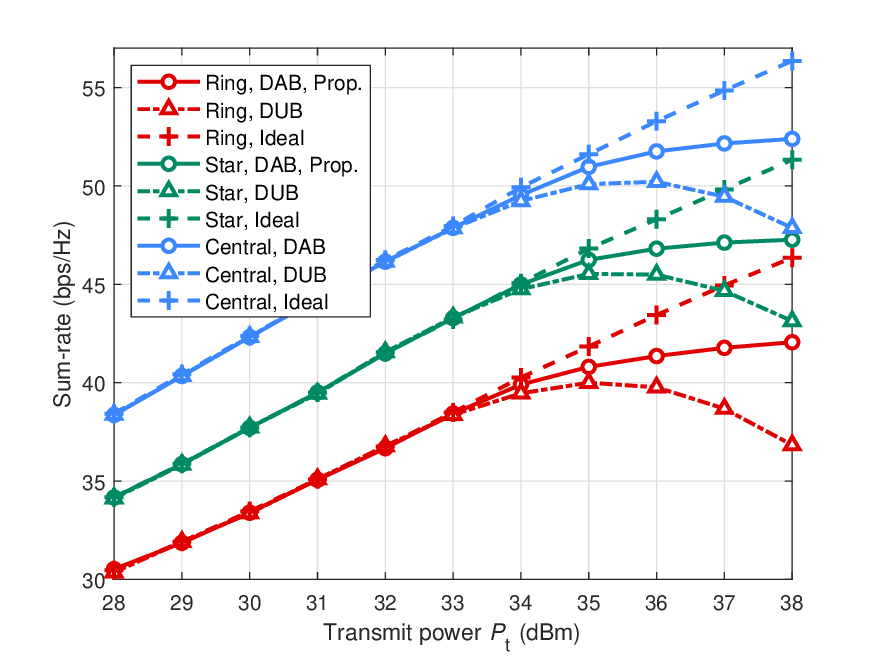}
  \caption{The sum-rate versus the transmit power $P_\text{t}$.}\label{fig:P}
\end{figure}

In the subsequent simulation results, we will evaluate the convergence, sum-rate, computational complexity, and overhead of the proposed distributed distortion-aware beamforming (DAB) designs, referred to as ``\textbf{Ring, DAB, Prop.}'' and ``\textbf{Star, DAB, Prop.}'' for the ring and star topologies, respectively. For comparison purposes, we also include the performance of the centralized distortion-aware beamforming design, denoted as ``\textbf{Central, DAB}".
To better highlight the performance gains of the distortion-aware beamforming design algorithms, we  compare them against several baseline schemes.  Specifically, the beamforming designs using ideal PAs are included as the performance upper-bounds, which are denoted as ``\textbf{Ring, Ideal}'', ``\textbf{Star, Ideal}'', and ``\textbf{Central, Ideal}'', respectively.
Additionally, the beamforming designs using practical PAs, without accounting for nonlinear amplification, are considered as performance lower bounds. These distortion-unaware beamforming (DUB) designs are are labeled as ``\textbf{Ring, DUB}'', ``\textbf{Star, DUB}'', and ``\textbf{Central, DUB}'', respectively.


The convergence performance of the proposed algorithms is illustrated in Fig. \ref{fig:convergence}. It is evident that all the proposed distortion-aware beamforming design algorithms exhibit fast convergence within a maximum of 15 iterations. And as the system scale grows with more BSs, the convergence speed of the algorithms gradually slows down due to the increased need for information exchange between the BSs. Similarly, the increase in the number of UEs enlarges the problem size, which results in slower convergence speed. Further comparisons reveal that the centralized algorithm converges more slowly than the distributed algorithms, primarily due to the greater computational difficulty associated with the high-dimensional joint beamforming design across all BSs, which hinders the convergence rate. Among the distributed approaches, the star-topology-based algorithm converges faster than the ring-topology-based algorithm, thanks to its advantage of information fusion at the central processor after local computations.

\begin{figure}[t]
  \centering
  \includegraphics[width= 3.1 in]{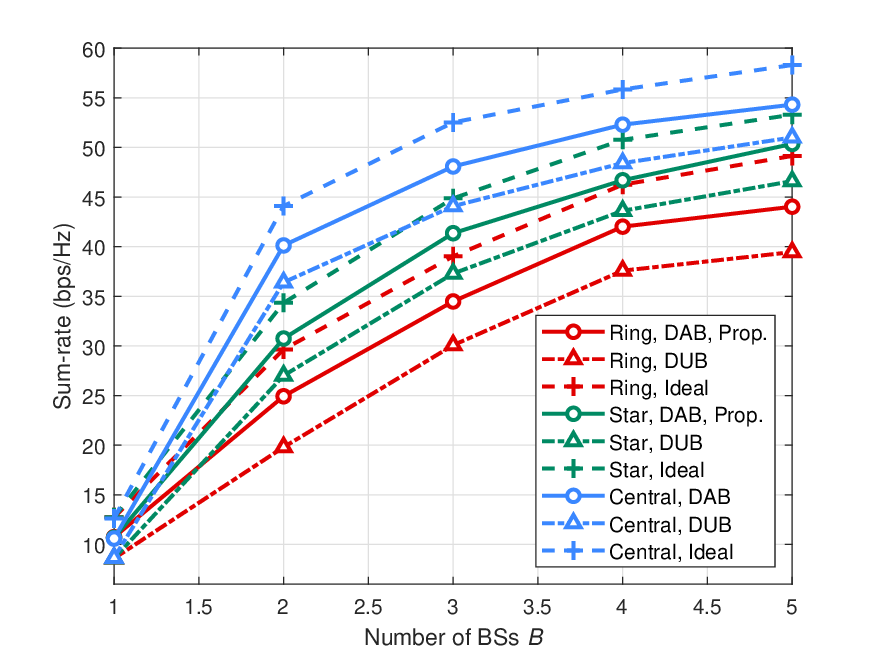}
  \caption{The sum-rate versus the number of BSs $B$.}\label{fig:B}
\end{figure}

\begin{figure}[t]
  \centering
  \includegraphics[width= 3.1 in]{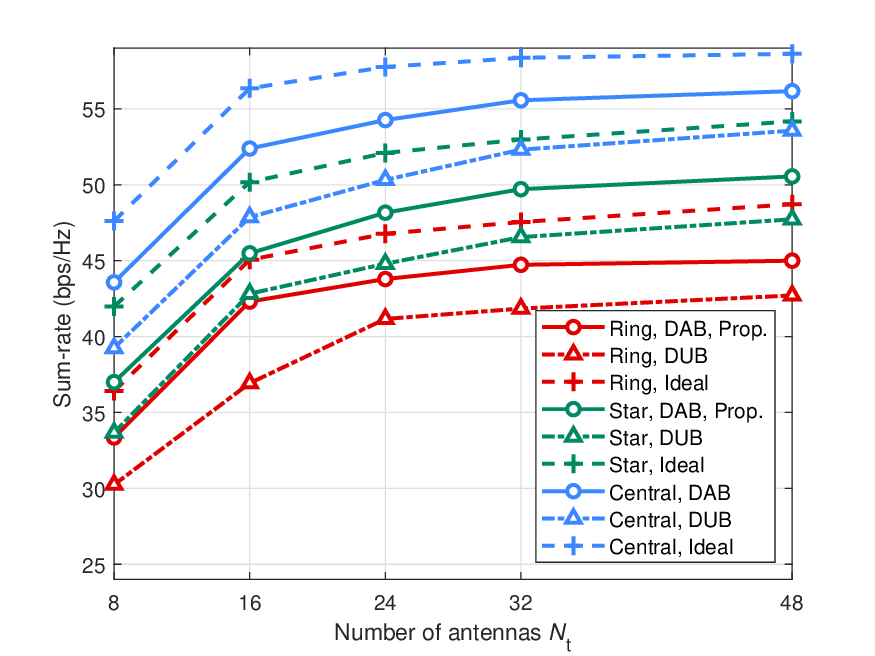}
  \caption{The sum-rate versus the number of antennas $N_{\mathrm{t}}$.}\label{fig:N}
\end{figure}

Fig. \ref{fig:P} illustrates the average sum-rate performance versus transmit power. The proposed distortion-aware beamforming designs achieve superior performance compared to algorithms that do not account for nonlinear distortion.
As anticipated, centralized algorithms have the best performance, followed by distributed algorithms.
In addition, benefiting from the data fusion ability, the star-topology-based algorithms outperform the ring-topology-based algorithms.
Interestingly, the sum-rates of distortion-unaware beamforming designs do not exhibit a monotonic increase with transmit power. This behavior can be explained by analyzing the SINDR expression in \eqref{eq:SINR_concise}. As the transmit power increases, both the useful signal power and interference power increase in a  proportional manner, while the distortion power increases at a faster rate as it exhibits greater sensitivity to the transmit power.
In the high-transmit-power regime, the nonlinear distortion becomes the dominant factor, overshadowing both the useful signal and interference. Consequently, further increases in transmit power lead to performance degradation as the nonlinear distortion is not effectively managed by distortion-unaware beamforming designs. This further highlights the importance of using our proposed distortion-aware beamforming algorithms to mitigate the adverse impact of nonlinear distortion on the system performance. Specifically, for $P_\text{t} = 38$dBm, the proposed distributed distortion-aware beamforming design algorithms are able to achieve approximately 90\% of the performance compared to the centralized approaches and 115\% of the performance compared to the distortion-unaware designs. It is also observed that the sum-rate of distortion-aware beamforming designs tends to saturate as the transmit power increases. Therefore, selecting an appropriate transmit power is crucial for achieving optimal energy efficiency while ensuring satisfactory system performance.

Next, the influence of the number of BSs on the average sum-rate is shown in Fig. \ref{fig:B} with $P_\text{t} = 38$ dBm. Obviously, an increase in the number of BSs provides more total power and spatial diversity, thereby resulting in higher sum-rates. However, as more BSs also introduce considerable distortion, the proposed distortion-aware beamforming has the potential to deliver a more significant performance enhancement. Similarly, Fig. \ref{fig:N} illustrates the impact of increasing the number of antennas on the sum-rate performance. As expected, a larger number of antennas with higher beamforming gain will deliver enhanced system capacity. Meanwhile, in larger antenna array scenarios with intensified distortion, our proposed distortion-aware beamforming shows greater promise in mitigating this effect and improving performance.

\begin{figure}[t]
    \centering
    \subfigure[]{\includegraphics[width= 1.5 in]{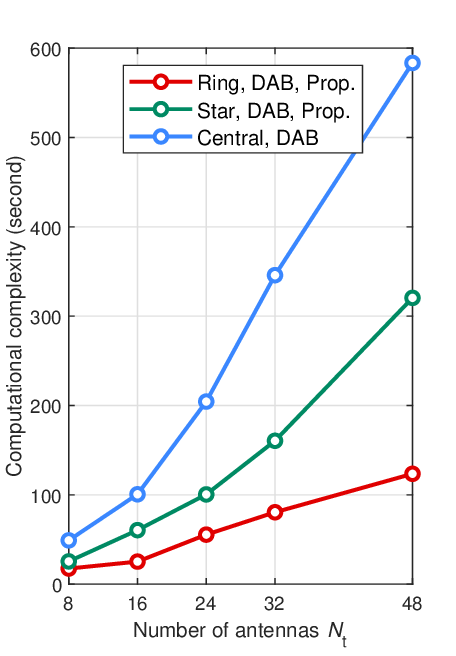}}%
    \vspace{-0.0 cm}
    \subfigure[]{\includegraphics[width= 1.5 in]{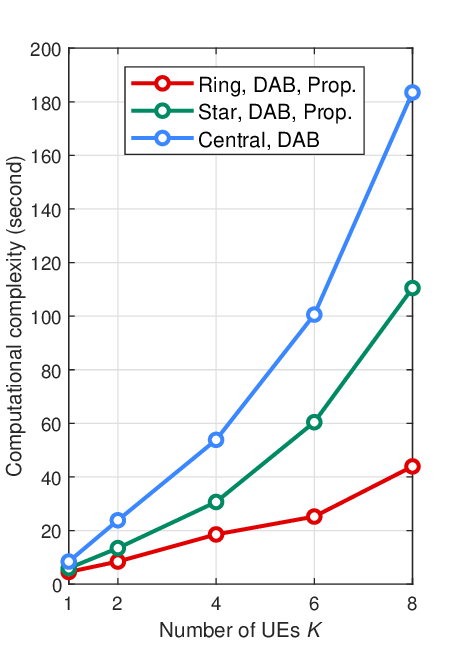}}
\caption{Comparison of computational complexity evaluated by algorithm convergence time.}
    \label{fig:complexity}
\end{figure}

\begin{figure}[t]
    \centering
    \subfigure[]{\includegraphics[width= 1.5 in]{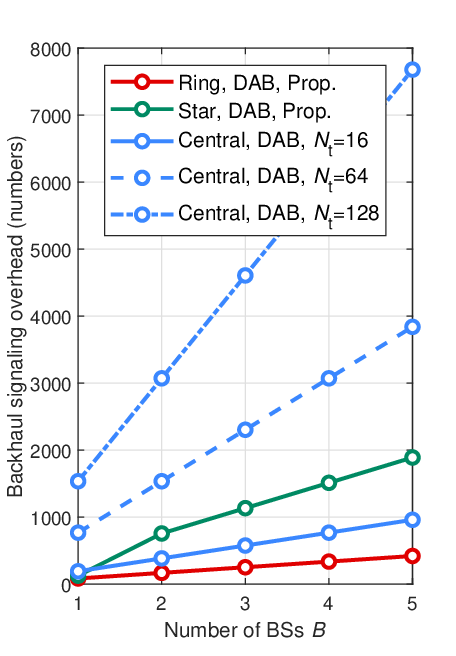}}%
    \vspace{-0.0 cm}
    \subfigure[]{\includegraphics[width= 1.5 in]{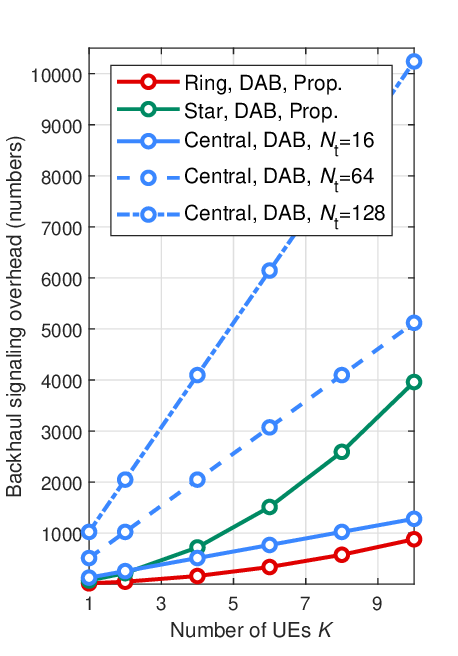}}
    \caption{Comparison of backhaul signaling overhead evaluated by the total amount of transferred double-precision numbers.}
    \label{fig:overhead}
\end{figure}

The comparisons of the proposed distortion-aware beamforming with respect to computational complexity and backhaul signaling overhead are illustrated in Fig. \ref{fig:complexity} and Fig. \ref{fig:overhead}, respectively.
We observe that the ring-topology-based design achieves a substantial reduction in computational complexity (about 80\%) and overhead (about 60\%), with only a 20\% performance loss compared to the centralized approach. In contrast, the star-topology-based design incurs relatively higher computational complexity and its overhead exceeds that of the centralized approach with $N_{\mathrm{t}}=16$. However, it is worth noting that the overhead of the distributed designs is independent of the number of antennas. Thus, the star-topology-based design remains promising, especially in the context of the growing trend toward extremely large-scale antenna arrays (ELAA) in future sixth-generation (6G) systems.


\begin{figure}[t]
    \centering
    \vspace{0.3 cm}
    \subfigure[BS-1]{\includegraphics[width= 2 in]{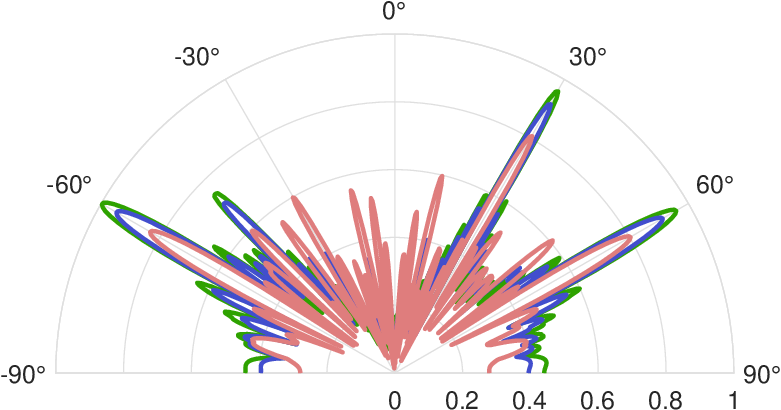}}
    \vspace{-0.0 cm}
       \subfigure[BS-3]{\includegraphics[width= 2 in]{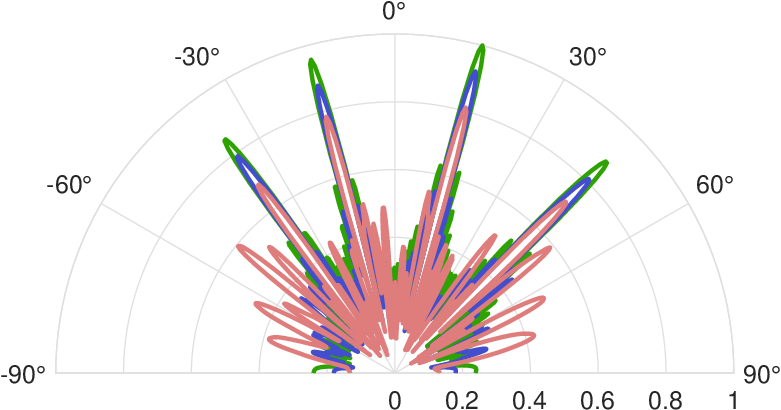}}
    \vspace{-0.0 cm}
    \caption{Illustration of the radiation patterns based on the ring topology. Green: Ideal; Blue: DAB; Red: DUB.}
    \label{fig:beampattern}
\end{figure}

Finally, we present the beam patterns aiming to intuitively reveal the impact of nonlinear distortion and demonstrate the effectiveness of the proposed distortion-aware beamforming algorithms.
To ensure clarity, we illustrate only the beam patterns of BS-1 and BS-3, using the same simulation setup as in Fig. \ref{fig:setting}. The system assumes $K=4$ UEs and $N_{\mathrm{t}}=64$ antennas. The proposed ring-topology-based distributed beamforming design is adopted as a distortion-aware beamforming solution owing to a more favorable trade-off among the performance, complexity, and overhead.
Specifically, the beam patterns are generated by ideal beamforming (green),   DUB design (red), and proposed DAB design  (blue), and illustrated in Fig. \ref{fig:beampattern} (a) and (b) for BS-1 and BS-3, respectively.
For DUB patterns which do not account for the effect of nonlinear PAs, we can clearly observe lower mainlobes and much larger sidelobes, indicating severe beam dispersion and distortion. In contrast, the proposed DAB design effectively enhances the mainlobes and suppresses the sidelobes, resulting in a pattern that closely resembles the ideal beamforming performance.

\section{Conclusions}
In this paper, we investigated the distortion-aware beamforming designs for CF-mMIMO systems, where the spatial distortion and beam dispersion caused by nonlinear PAs were considered. Based on the typical third-order memoryless polynomial distortion model, we proposed distortion-aware beamforming design to maximize the achievable sum-rate of the CF-mMIMO system under the transmit power constraint of each BS. This complicated distortion-aware beamforming design problem was solved in a distributed manner based on the ring and star topologies, respectively.
Extensive simulation results demonstrated the substantial performance improvement achieved by utilizing the proposed distributed distortion-aware beamforming designs to mitigate nonlinear PA distortion.
As imperfect channel estimation also poses significant challenges in real-world wireless communication scenarios, our future research will focus on developing a robust beamforming strategy to mitigate the combined effects of distortion and imperfect CSI.

\begin{appendices}
\section{}
The minimum of the convex objective function is attained when its derivative with respect to $\mathbf{R}_{b}$ equals zero. To facilitate the analysis, the objective function is divided into following four components
\begin{equation}
\begin{aligned}
\!\!\!f_{1}(\mathbf{R}_{b})&=\mathbf{w}_{b}^{H}\mathbf{\overline{G}}_{b}^{H}\mathbf{E}_{3,b}\mathbf{\overline{G}}_{b}\mathbf{w}_{b}=\mathrm{Tr}(\mathbf{\overline{G}}_{b}\mathbf{w}_{b}\mathbf{w}_{b}^{H}\mathbf{\overline{G}}_{b}^{H}\mathbf{E}_{3,b}),\\
\!\!\!f_{2}(\mathbf{R}_{b})&=\Re\{(2\mathbf{e}_{b}^{H}\mathbf{\overline{G}}_{b}-\mathbf{h}_{b}^{H}\mathbf{E}_{2}\mathbf{\overline{G}}_{b})\mathbf{w}_{b}\},\\
\!\!\!f_{3}(\mathbf{R}_{b})&=\mathbf{h}_{b}^{H}\mathbf{E}_{4}^{H}\mathbf{\overline{C}}_{\mathrm{d},b}\mathbf{E}_{4}\mathbf{h}_{b},\\
\!\!\!f_{4}(\mathbf{R}_{b})&=\rho\|\mathbf{R}_{b}-\mathbf{w}_{b}\mathbf{w}_{b}^H\|_{F}^{2}.
\end{aligned}
\end{equation}
Then, the derivative of each component with respect to $\mathbf{R}_{b}$ is computed individually by using a divide-and-conquer approach.
\newcounter{TempEqCnt3}
\setcounter{TempEqCnt3}{\value{equation}}
\setcounter{equation}{72}
\begin{figure*}[b]
\hrulefill
\begin{equation}\label{eq:vec_Rb}
\small
\begin{aligned}
\!\!&\ \ \ \ \mathrm{vec}(\mathbf{E}_{3,b}^{T}\mathbf{\overline{G}}_{b}^{\ast}(\mathbf{w}_{b}\mathbf{w}_{b}^{H})^{T})\\
\!\!&=((\mathbf{w}_{b}\mathbf{w}_{b}^{H})\otimes \mathbf{E}_{3,b}^{T})\mathrm{vec}(\mathbf{I}_{K}\otimes(\beta_{1}^{\ast}\mathbf{I}_{N_{\mathrm{t}}}
\!+\!2\beta_{3}^{\ast}\mathbf{E}_{1}^{T}(\mathbf{R}_{b}^{\ast}\odot\mathbf{I}_{N_{\mathrm{t}}K})\mathbf{E}_{1}))\\
\!\!&=((\mathbf{w}_{b}\mathbf{w}_{b}^{H})\otimes \mathbf{E}_{3,b}^{T})\mathrm{vec}(\mathbf{I}_{K}\otimes(\beta_{1}^{\ast}\mathbf{I}_{N_{\mathrm{t}}}))
\!+\!((\mathbf{w}_{b}\mathbf{w}_{b}^{H})\otimes \mathbf{E}_{3,b}^{T})\mathrm{vec}((\mathbf{E}_{1}(2\beta_{3}^{\ast}\mathbf{E}_{1}^{T}(\mathbf{R}_{b}^{\ast}\odot\mathbf{I}_{N_{\mathrm{t}}K})\mathbf{E}_{1})\mathbf{E}_{1}^{T})\odot\mathbf{I}_{N_{\mathrm{t}}K})\\
\!\!&=((\mathbf{w}_{b}\mathbf{w}_{b}^{H})\otimes \mathbf{E}_{3,b}^{T})\mathrm{vec}(\mathbf{I}_{K}\otimes(\beta_{1}^{\ast}\mathbf{I}_{N_{\mathrm{t}}}))
\!+\!2\beta_{3}^{\ast}((\mathbf{w}_{b}\mathbf{w}_{b}^{H})\otimes \mathbf{E}_{3,b}^{T})\mathrm{diag}\{\mathrm{vec}(\mathbf{I}_{N_{\mathrm{t}}K})\}((\mathbf{E}_{1}\mathbf{E}_{1}^{T})\otimes(\mathbf{E}_{1}\mathbf{E}_{1}^{T}))\mathrm{diag}\{\mathrm{vec}(\mathbf{I}_{N_{\mathrm{t}}K})\}\mathrm{vec}(\mathbf{R}_{b}^{\ast}).\\
\end{aligned}
\end{equation}
\setcounter{equation}{77}
\begin{equation}\label{eq:CRcR}
\small
\begin{aligned}
\mathbf{C}_{\mathrm{R},b}&=2\beta_{3}^{\ast}(2\beta_{3}\mathbf{B}_{R}(\mathbf{E}_{1}^{T}\otimes\mathbf{E}_{1}^{T})\mathrm{diag}\{\mathrm{vec}(\mathbf{I}_{N_{\mathrm{t}}K})\})^{T} ((\mathbf{w}_{b}\mathbf{w}_{b}^{H})\otimes \mathbf{E}_{3,b}^{T})\mathrm{diag}\{\mathrm{vec}(\mathbf{I}_{N_{\mathrm{t}}K})\}((\mathbf{E}_{1}\mathbf{E}_{1}^{T})\otimes(\mathbf{E}_{1}\mathbf{E}_{1}^{T}))\mathrm{diag}\{\mathrm{vec}(\mathbf{I}_{N_{\mathrm{t}}K})\},\\
\mathbf{c}_{\mathrm{R},b}&=(2\beta_{3}\mathbf{B}_{R}(\mathbf{E}_{1}^{T}\otimes\mathbf{E}_{1}^{T})\mathrm{diag}\{\mathrm{vec}(\mathbf{I}_{N_{\mathrm{t}}K})\})^{T}((\mathbf{w}_{b}\mathbf{w}_{b}^{H})\otimes \mathbf{E}_{3,b}^{T})\mathrm{vec}(\mathbf{I}_{K}\otimes(\beta_{1}^{\ast}\mathbf{I}_{N_{\mathrm{t}}}))
\\
&\ \ \ \ +(2\beta_{3}\mathbf{B}_{R}(\mathbf{E}_{1}^{T}\otimes\mathbf{E}_{1}^{T})\mathrm{diag}\{\mathrm{vec}(\mathbf{I}_{N_{\mathrm{t}}K})\})^{T}
\mathrm{vec}\big(\frac{1}{2}(\mathbf{w}_{b}(2\mathbf{e}_{b}^{H}-\mathbf{h}_{b}^{H}\mathbf{E}_{2}))^{T}\big)
\\
&\ \ \ \ +(2|\beta_{3}|^{2}\mathbf{B}_{R}\mathrm{diag}\{\mathrm{vec}(|\mathbf{\overline{F}}_{b}^{(t)}|^{2})\}
(\mathbf{E}_{1}^{T}\otimes\mathbf{E}_{1}^{T})\mathrm{diag}\{\mathrm{vec}(\mathbf{I}_{K}\otimes\mathbf{1}_{N_{\mathrm{t}}\times N_{\mathrm{t}}})\})^{T} \mathrm{vec}\big((\mathbf{E}_{4}\mathbf{h}_{b}\mathbf{h}_{b}^{H}\mathbf{E}_{4}^{H})^{T}\big)-\rho\mathrm{vec}((\mathbf{w}_{b}\mathbf{w}_{b}^H)^{T}).
\end{aligned}
\end{equation}
\end{figure*}

\setcounter{equation}{69}
By taking the partial derivative $f_{1}(\mathbf{R}_{b})$ with respect to $\mathbf{R}_{b}$, we obtain\begin{equation}\label{eq:f1}
\begin{aligned}
\frac{\partial f_{1}(\mathbf{R}_{b})}{\partial\mathbf{R}_{b}}
&=\frac{\partial f_{1}(\mathbf{R}_{b})}{\partial\mathbf{\overline{G}}_{b}}\frac{\partial \mathbf{\overline{G}}_{b}}{\partial\mathbf{R}_{b}}+\frac{\partial f_{1}(\mathbf{R}_{b})}{\partial\mathbf{\overline{G}}_{b}^{\ast}}\frac{\partial \mathbf{\overline{G}}_{b}^{\ast}}{\partial\mathbf{R}_{b}}\\
&=\mathrm{vec}(\mathbf{E}_{3,b}^{T}\mathbf{\overline{G}}_{b}^{\ast}(\mathbf{w}_{b}\mathbf{w}_{b}^{H})^{T})^{T}\frac{\partial \mathbf{\overline{G}}_{b}}{\partial\mathbf{R}_{b}},
\end{aligned}
\end{equation}
where the derivative of the inner function, i.e., $\frac{\partial \mathbf{\overline{G}}_{b}}{\partial\mathbf{R}_{b}}$ is calculated as
\begin{equation}
\begin{aligned}
\frac{\partial \mathbf{\overline{G}}_{b}}{\partial\mathbf{R}_{b}}&=\mathbf{B}_{R}\frac{\partial \mathbf{G}_{b}}{\partial\mathbf{R}_{b}},\\
\frac{\partial \mathbf{G}_{b}}{\partial\mathbf{R}_{b}}&=2\beta_{3}(\mathbf{E}_{1}^{T}\otimes\mathbf{E}_{1}^{T})\mathrm{diag}\{\mathrm{vec}(\mathbf{I}_{N_{\mathrm{t}}K})\},
\end{aligned}
\end{equation}
\vspace{-0.2cm}
in which the constant matrix $\mathbf{B}_{R}$ can be determined by
\begin{equation}
\begin{aligned}
\mathbf{a}&\triangleq [1, 0 , \ldots, 0]^{T} \in \mathbb{R}^{N_{\mathrm{t}}+1},\mathbf{b}\triangleq [1, 0 , \ldots, 0]^{T} \in \mathbb{R}^{N_{\mathrm{t}}K+1},\\
\mathbf{A}_{1}&\triangleq \mathbf{I}_{N_{\mathrm{t}}}\otimes \mathbf{a}^{T} \in \mathbb{R}^{N_{\mathrm{t}}\times (N_{\mathrm{t}}^{2}+N_{\mathrm{t}})},\\
\mathbf{A}_{2}&\triangleq [\mathbf{I}_{N_{\mathrm{t}}^{2}}, \mathbf{0}_{N_{\mathrm{t}}^{2}\times N_{\mathrm{t}}}]^{T} \in \mathbb{R}^{(N_{\mathrm{t}}^{2}+N_{\mathrm{t}})\times N_{\mathrm{t}}^{2}},\\
\mathbf{A}_{R}&\triangleq \mathbf{A}_{1}\mathbf{A}_{2}\in \mathbb{R}^{N_{\mathrm{t}}\times N_{\mathrm{t}}^{2}},
\mathbf{B}_{1}\triangleq \mathbf{A}_{R}\otimes\mathbf{b}\in \mathbb{R}^{(N_{\mathrm{t}}^{2}K+N_{\mathrm{t}})\times N_{\mathrm{t}}^{2}},\\
\mathbf{B}_{2}&\triangleq\mathbf{1}_{K}\otimes \mathbf{B}_{1}\in \mathbb{R}^{(N_{\mathrm{t}}^{2}K^{2}+N_{\mathrm{t}}K)\times N_{\mathrm{t}}^{2}},\\
\mathbf{B}_{3}&\triangleq[\mathbf{I}_{N_{\mathrm{t}}^{2}K^{2}}, \mathbf{0}_{N_{\mathrm{t}}^{2}K^{2}\times N_{\mathrm{t}}K}]\in \mathbb{R}^{N_{\mathrm{t}}^{2}K^{2}\times (N_{\mathrm{t}}^{2}K^{2}+N_{\mathrm{t}}K)},\\
\mathbf{B}_{R}&\triangleq\mathbf{B}_{3}\mathbf{B}_{2}\in \mathbb{R}^{N_{\mathrm{t}}^{2}K^{2}\times N_{\mathrm{t}}^{2}}.
\end{aligned}
\end{equation}
Additionally, to facilitate deriving the closed-form solution for variable $\mathbf{R}_{b}$ afterward, we can reformulate the term $\mathrm{vec}(\mathbf{E}_{3,b}^{T}\mathbf{\overline{G}}_{b}^{\ast}(\mathbf{w}_{b}\mathbf{w}_{b}^{H})^{T})$ in \eqref{eq:f1} into an explicit expression for $\mathbf{R}_{b}$ as shown in \eqref{eq:vec_Rb} at the bottom of this page.

By applying the similar method, the partial derivatives of $f_{2}(\mathbf{R}_{b})$ and $f_{3}(\mathbf{R}_{b})$ with respect to $\mathbf{R}_{b}$ can be derived as
\setcounter{equation}{73}
\begin{equation}
\begin{aligned}
\frac{\partial f_{2}(\mathbf{R}_{b})}{\partial\mathbf{R}_{b}}
&=\mathrm{vec}\big(\frac{1}{2}(\mathbf{w}_{b}(2\mathbf{e}_{b}^{H}-\mathbf{h}_{b}^{H}\mathbf{E}_{2}))^{T}\big)^{T}\frac{\partial \mathbf{\overline{G}}_{b}}{\partial\mathbf{R}_{b}},\\
\frac{\partial f_{3}(\mathbf{R}_{b})}{\partial\mathbf{R}_{b}}
&=\mathrm{vec}\big((\mathbf{E}_{4}\mathbf{h}_{b}\mathbf{h}_{b}^{H}\mathbf{E}_{4}^{H})^{T}\big)^{T}\frac{\partial \mathbf{\overline{C}}_{\mathrm{d},b}}{\partial\mathbf{R}_{b}},
\end{aligned}
\end{equation}
where the derivative of the inner function concerning $\mathbf{R}_{b}$, i.e., $\frac{\partial \mathbf{\overline{C}}_{\mathrm{d},b}}{\partial\mathbf{R}_{b}}$, is determined as
\begin{equation}
\begin{aligned}
\frac{\partial \mathbf{\overline{C}}_{\mathrm{d},b}}{\partial\mathbf{R}_{b}} =  &
 \mathbf{B}_{R}\frac{\partial \mathbf{C}_{\mathrm{d},b}}{\partial\mathbf{R}_{b}},\\
\frac{\partial \mathbf{C}_{\mathrm{d},b}}{\partial\mathbf{R}_{b}} = & 2|\beta_{3} |^{2} \times
\mathrm{diag}\{\mathrm{vec}(|\mathbf{\overline{F}}_{b}^{(t)} |^{2})\}
(\mathbf{E}_{1}^{T} \otimes \mathbf{E}_{1}^{T} ) \times \\ &  \mathrm{diag}\{ \mathrm{vec}(\mathbf{I}_{K} \otimes \mathbf{1}_{N_{\mathrm{t}}\!\times\! N_{\mathrm{t}}} ) \}.
\end{aligned}
\end{equation}
And the partial derivative of $f_{4}(\mathbf{R}_{b})$ with respect to $\mathbf{R}_{b}$ is calculated as
\begin{equation}
\frac{\partial f_{4}(\mathbf{R}_{b})}{\partial\mathbf{R}_{b}}
=\rho\mathrm{vec}(\mathbf{R}_{b}^{\ast}-(\mathbf{w}_{b}\mathbf{w}_{b}^H)^{T})^{T}.
\end{equation}

Finally, we can set the first-order derivative of the original objective function in \eqref{eq:Problem_formulation_opt_R} with respect to $\mathbf{R}_{b}$ to zero to obtain the closed-form solution, which yields
\begin{equation}
\mathbf{C}_{\mathrm{R},b}\mathrm{vec}(\mathbf{R}_{b}^{\ast})+\mathbf{c}_{\mathrm{R},b}+\rho\mathrm{vec}(\mathbf{R}_{b}^{\ast})=\mathbf{0},
\end{equation}
where the coefficient matrix $\mathbf{C}_{\mathrm{R},b}$ and vector $\mathbf{c}_{\mathrm{R},b}$ are formulated in \eqref{eq:CRcR} at the bottom of previous page.
Consequently, the closed-form solution for $\mathbf{R}_{b}$ in solving convex problem \eqref{eq:Problem_formulation_opt_R} can be expressed as:
\setcounter{equation}{78}
\begin{equation}
\mathrm{vec}(\mathbf{R}_{b})=-((\mathbf{C}_{\mathrm{R},b}+\rho\mathbf{I}_{N_{\mathrm{t}}^{2}K^{2}})^{-1}\mathbf{c}_{\mathrm{R},b})^{\ast}.
\end{equation}
\end{appendices}


\begin{thebibliography}{99}

\bibitem{E. Bjornson 2019 MIMO} E. Bj\"{o}rnson, L. Sanguinetti, H. Wymeersch, J. Hoydis, and T. L. Marzetta, ``Massive MIMO is a reality-what is next?: Five promising research directions for antenna arrays," \textit{Digit. Signal Process.}, vol. 94, pp. 3-20, Nov. 2019.


\bibitem{E. Bjornson 2014}E. Bj\"{o}rnson, J. Hoydis, M. Kountouris, and M. Debbah, ``Massive MIMO systems with non-ideal hardware: Energy efficiency, estimation, and capacity limits," \textit{IEEE Trans. Inf. Theory}, vol. 60, no. 11, pp. 7112-7139, Nov. 2014.

\bibitem{A. L. Swindlehurst TWC 2023} Z. Peng, R. Weng, C. Pan, G. Zhou, M. D. Renzo, and A. L. Swindlehurst, ``Robust transmission design for RIS-assisted secure multiuser communication systems in the presence of hardware impairments," \textit{IEEE Trans. Wireless Commun.}, vol. 22, no. 11, pp. 7506-7512, Nov. 2023.

\bibitem{E. Jorswieck 2023} M. Soleymani, I. Santamaria, and E. Jorswieck, ``NOMA-based improper signaling for MIMO STAR-RIS-assisted broadcast channels with hardware impairments," in \textit{Proc. IEEE Global Commun. Conf. (GLOBECOM)}, Kuala Lumpur, Malaysia, Dec. 2023.

\bibitem{E. G. Larsson OJCS 2024} U. K. Ganesan, T. T. Vu, and E. G. Larsson, ``Cell-Free massive MIMO with multi-antenna users and phase misalignments: A novel partially coherent transmission framework," \textit{IEEE Open J. Commun. Soc.}, vol. 5, pp. 1639-1655, Mar. 2024.



\bibitem{X. Li PIMRC 2023} T. Du, J. Yang, X. Yi, X. Li, and S. Jin, ``Reciprocity calibration for massive MIMO with low-resolution ADCs," in \textit{Proc. IEEE Int. Symp. Personal, Indoor Mobile Radio Commun. (PIMRC)}, Toronto, ON, Canada, Oct. 2023.

\bibitem{A. L. Swindlehurst TWC 2021} L. V. Nguyen, A. L. Swindlehurst, and D. H. N. Nguyen, ``Linear and deep neural network-based receivers for massive MIMO systems with one-bit ADCs," \textit{IEEE Trans. Wireless Commun.}, vol. 20, no. 11, pp. 7333-7345, Nov. 2021.

\bibitem{C. Mollen 2018} C. Moll\'{e}n, U. Gustavsson, T. Eriksson, and E. G. Larsson, ``Spatial characteristics of distortion radiated from antenna arrays with transceiver nonlinearities," \textit{IEEE Trans. Wireless Commun.}, vol. 17, no. 10, pp. 6663-6679, Oct. 2018.

\bibitem{N. N. Moghadam 2018} N. N. Moghadam, G\'{a}bor Fodor, M. Bengtsson, and D. J. Love, ``On the energy efficiency of MIMO hybrid beamforming for millimeter-wave systems with nonlinear power amplifiers," \textit{IEEE Trans. Wireless Commun.}, vol. 17, no. 11, pp. 7208-7221, Nov. 2018.

\bibitem{E. Bjornson 2024} N. Kolomvakis, A. Kosasih, and E. Bj\"{o}rnson, ``Nonlinear distortion radiated from large arrays and active reconfigurable intelligent surfaces," Jan. 2024. [Online]. Available: https://arxiv.org/abs/2401.12622

\bibitem{E. Bjornson 2019} E. Bj\"{o}rnson, L. Sanguinetti, and J. Hoydis, ``Hardware distortion correlation has negligible impact on UL massive MIMO spectral efficiency," \textit{IEEE Trans. Commun.}, vol. 67, no. 2, pp. 1085-1098, Feb. 2019.

\bibitem{B. Liu 2024} B. Liu, Q. Wang, and S. Pollin, ``TAIS: Transparent amplifying intelligent surface for indoor-to-outdoor mmWave communications," \textit{IEEE Trans. Commun.}, vol. 72, no. 2, pp. 1223-1238, Feb. 2019.

\bibitem{S. R. Aghdam 2019} S. R. Aghdam, S. Jacobsson, and T. Eriksson, ``Distortion-aware linear precoding for millimeter-wave multiuser MISO downlink," in \textit{Proc. IEEE Int. Conf. Commun. Workshops}, Shanghai, China, May 2019.

\bibitem{M. Wu 2022} M. Wu, M. Li, M. Zhao, and M. Zhao, ``A WMMSE approach to distortion-aware beamforming design for millimeter-wave massive MIMO downlink communication," in \textit{Proc. IEEE 95th Veh. Technol. Conf. (VTC)}, Helsinki, Finland, Aug. 2022.


\bibitem{JSTSP} Y. Xu, E. G. Larsson, E. A. Jorswieck, X. Li, S. Jin, and T.-H. Chang, ``Distributed signal processing for extremely large-scale antenna array systems: State-of-the-art and future directions," \textit{IEEE J. Sel. Topics Signal Process. (JSTSP)}, early access, Feb. 2025. doi: 10.1109/JSTSP.2025.3541386.

\bibitem{W. Yu decentralized 2024} Y.-F. Liu, T.-H. Chang, M. Hong, Z. Wu, A. M.-C. So, E. A. Jorswieck, and W. Yu, ``A survey of recent advances in optimization methods for wireless communications," \textit{IEEE J. Sel. Areas Commun.}, vol. 42, no. 11, pp. 2992-3031, Aug. 2024.

\bibitem{T.-H. Chang ICC 2023} T. Cai, S. Ge, Y. Xu, and T.-H. Chang, ``Approaching centralized multi-cell coordinated beamforming with limited backhaul signaling," in \textit{Proc. IEEE Int. Conf. Commun. (ICC)}, Rome, Italy, Oct. 2023.

\bibitem{T.-H. Chang ICASSP 2023} Y. Xu, E. Song, Q. Shi, and T.-H. Chang, ``Sparse aggregation-based channel estimation for massive MIMO systems with decentralized baseband processing," in \textit{Proc. IEEE Int. Conf. Acoust. Speech Signal Process. (ICASSP)}, Rhodes Island, Greece, Jun. 2023.

\bibitem{T.-H. Chang J-1 2023} Y. Xu, B. Wang, E. Song, Q. Shi, and T.-H. Chang, ``Low-complexity channel estimation for massive MIMO systems with decentralized baseband processing," \textit{IEEE Trans. Signal Process.}, vol. 71, pp. 2728-2743, Jul. 2023.

\bibitem{T.-H. Chang J-2 2023} X. Zhao, M. Li, Y. Liu, T.-H. Chang, and Q. Shi, ``Communication-efficient decentralized linear precoding for massive MU-MIMO systems," \textit{IEEE Trans. Signal Process.}, vol. 71, pp. 4045-4059, Oct. 2023.

\bibitem{E. G. Larsson WCL-2 2024} Z. H. Shaik and E. G. Larsson, ``Decentralized algorithms for out-of-system interference suppression in distributed MIMO," \textit{IEEE Wireless Commun. Lett.}, vol. 13, no. 7, pp. 1953-1957, Jul. 2024.

\bibitem{E. Jorswieck 2016} A. Zappone, E. Jorswieck, and A. Leshem, ``Distributed resource allocation for energy efficiency in MIMO OFDMA wireless networks," \textit{IEEE J. Sel. Areas Commun.}, vol. 34, no. 12, pp. 3451-3465, Oct. 2016.

\bibitem{H. A. Ammar TWC 2018}  H. A. Ammar, R. Adve, S. Shahbazpanahi, G. Boudreau, and K. V. Srinivas, ``Distributed resource allocation optimization for user-centric cell-rree MIMO networks," \textit{IEEE Trans. Wireless Commun.}, vol. 8, no. 3, pp. 3099-3115, May 2022.

\bibitem{H. Vincent Poor 2021} S. Huang, Y. Ye, M. Xiao, H. Vincent Poor, and M. Skoglund, ``Decentralized beamforming design for intelligent reflecting surface-enhanced cell-free networks," \textit{IEEE Wireless Commun. Lett.}, vol. 10, no. 3, pp. 673-677, Mar. 2021.




\bibitem{A. Tolli 2011} A. T\"{o}lli, H. Pennanen, and P. Komulainen, ``Decentralized minimum power multi-cell beamforming with limited backhaul signaling," \textit{IEEE Trans. Wireless Commun.}, vol. 8, no. 3, pp. 570-580, Feb. 2011.


\bibitem{T.-H. Chang J-1 2020} Q. Shi and M. Hong, ``Penalty dual decomposition method for nonsmooth nonconvex optimization-part I: Algorithms and convergence analysis," \textit{IEEE Trans. Signal Process.}, vol. 68, pp. 4108-4122, Jun. 2020.

\bibitem{T.-H. Chang J-2 2020} Q. Shi, M. Hong, X. Fu, and T.-H. Chang, ``Penalty dual decomposition method for nonsmooth nonconvex optimization-part II: Applications," \textit{IEEE Trans. Signal Process.}, vol. 68, pp. 4242-4257, Jun. 2020.

\bibitem{ADMM 2011} S. Boyd, N. Parikh, E. Chu, B. Peleato, and J. Eckstein, ``Distributed optimization and statistical learning via the alternating direction method of multipliers," \textit{Found. Trends Mach. Learn.}, vol. 3, no. 1, pp. 1-122, Jan. 2011.


\bibitem{T.-H. Chang 2016-J1 consensus ADMM } T.-H. Chang, M. Hong, W.-C. Liao, and X. Wang, ``Asynchronous distributed ADMM for large-scale optimization-part I: Algorithm and convergence analysis," \textit{IEEE Trans. Signal Process.}, vol. 64, no. 12, pp. 3118-3130, Jun. 2016.


\bibitem{T.-H. Chang 2016-J2 consensus ADMM} T.-H. Chang, M. Hong, W.-C. Liao, and X. Wang, ``Asynchronous distributed ADMM for large-scale optimization-part II: Linear convergence analysis and numerical performance," \textit{IEEE Trans. Signal Process.}, vol. 64, no. 12, pp. 3131-3144, Jun. 2016.


\bibitem{T.-H. Chang 2016-J3 consensus ADMM} T.-H. Chang, ``A proximal dual consensus ADMM method for multi-agent constrained optimization," \textit{IEEE Trans. Signal Process.}, vol. 64, no. 14, pp. 3719-3734, Jun. 2016.


\bibitem{T.-H. Chang 2012 consensus ADMM}  C. Shen, T.-H. Chang, K.-Y. Wang, Z. Qiu, and C.-Y. Chi, ``Distributed robust multicell coordinated beamforming with imperfect CSI: An ADMM approach," \textit{IEEE Trans. Signal Process.}, vol. 60, no. 6, pp. 2988-3003, Jun. 2012.


\bibitem{M. Maros 2018 consensus ADMM}  M. Maros and J. Jald\'{e}n, ``ADMM for distributed dynamic beamforming," \textit{IEEE Trans. Signal Inf. Process. over Netw.}, vol. 4, no. 2, pp. 220-235, Jun. 2018.


\bibitem{F. Han ADMM 2017} F. Han, S. Zhao, L. Zhang, K. Yang, and J. Shen, ``Decentralized beamforming for weighted sum energy efficiency maximization in MIMO systems," in \textit{Proc. IEEE Global Commun. Conf. (GLOBECOM)}, Singapore, Dec. 2017.

\bibitem{P. Ni decentralized 2023} P. Ni, M. Li, R. Liu, and Q. Liu, ``Partially distributed beamforming design for RIS-aided cell-free networks," \textit{IEEE Trans. Veh. Technol.}, vol. 71, no. 12, pp. 13377-13381, Dec. 2022.


\bibitem{H. Zhang decentralized 2023} H. Zhang, S. Liu, R. Liu, M. Li, and Q. Liu, ``Distributed DRL based beamforming design for RIS-assisted multi-cell systems," in \textit{Proc. IEEE Global Commun. Conf. (GLOBECOM)}, Kuala Lumpur, Malaysia, Dec. 2023.

\bibitem{Z. Wang decentralized 2022} Z. Wang, M. Eisen, and A. Ribeiro, ``Learning decentralized wireless resource allocations with graph neural networks," \textit{IEEE Trans. Signal Process.}, vol. 70, pp. 1850-1863, Mar. 2022.

%
\bibitem{E. Bjornson 2021} \"{O}. T. Demir and E. Bjornson, ``The Bussgang decomposition of nonlinear systems: Basic theory and MIMO extensions," \textit{IEEE Signal Process. Mag.}, vol. 38, no. 1, pp. 131-136, Jan. 2021.

\bibitem{FP} K. Shen and W. Yu, ``Fractional programming for communication systems-Part II: Uplink scheduling via matching," \textit{IEEE Trans. Signal Process.}, vol. 66, no. 10, pp. 2631-2644, May 2018.

\bibitem{H. Li 2020 Hybrid} H. Li, M. Li, and Q. Liu, ``Hybrid beamforming with dynamic subarrays and low-resolution PSs for mmWave MU-MISO systems," \textit{IEEE Trans. Commun.}, vol. 68, no. 1, pp. 602-614, Jan. 2020.

\bibitem{MM} Y. Sun, P. Babu, and D. P. Palomar, ``Majorization-minimization algorithms in signal processing, communications, and machine learning," \textit{IEEE Trans. Signal Process.}, vol. 65, no. 3, pp. 794-816, Feb. 2017.

\bibitem{R. Liu} R. Liu, M. Li, Y. Liu, Q. Wu, and Q. Liu, ``Joint transmit waveform and passive beamforming design for RIS-aided DFRC systems," \textit{IEEE J. Sel. Topics Signal Process. (JSTSP)}, vol. 16, no. 5, pp. 995-1010, Aug. 2022.



\end{thebibliography}
\end{document}